\documentclass[twoside,twocolumn,9pt]{article}
\usepackage{extsizes}
\usepackage[super,sort&compress,comma]{natbib} 
\usepackage[version=3]{mhchem}
\usepackage[left=1.5cm, right=1.5cm, top=1.785cm, bottom=2.0cm]{geometry}
\usepackage{balance}
\usepackage{mathptmx}
\usepackage{sectsty}
\usepackage{graphicx} 
\usepackage{lastpage}
\usepackage[format=plain,justification=justified,singlelinecheck=false,font={stretch=1.125,small,sf},labelfont=bf,labelsep=space]{caption}
\usepackage{float}
\usepackage{fancyhdr}
\usepackage{fnpos}
\usepackage[english]{babel}
\addto{\captionsenglish}{%
  
}
\usepackage{array}
\usepackage{droidsans}
\usepackage{charter}
\usepackage[T1]{fontenc}
\usepackage[usenames,dvipsnames]{xcolor}
\usepackage{setspace}
\usepackage[compact]{titlesec}
\usepackage{hyperref}
\usepackage{ulem}
\usepackage{amssymb}
\usepackage{verbatim}

\usepackage{epstopdf}

\definecolor{cream}{RGB}{222,217,201}

\begin{document}

\pagestyle{fancy}
\thispagestyle{plain}
\fancypagestyle{plain}{
\renewcommand{\headrulewidth}{0pt}
}

\makeFNbottom
\makeatletter
\renewcommand\LARGE{\@setfontsize\LARGE{15pt}{17}}
\renewcommand\Large{\@setfontsize\Large{12pt}{14}}
\renewcommand\large{\@setfontsize\large{10pt}{12}}
\renewcommand\footnotesize{\@setfontsize\footnotesize{7pt}{10}}
\makeatother

\renewcommand{\thefootnote}{\fnsymbol{footnote}}
\renewcommand\footnoterule{\vspace*{1pt}%
\color{cream}\hrule width 3.5in height 0.4pt \color{black}\vspace*{5pt}} 
\setcounter{secnumdepth}{5}

\makeatletter 
\renewcommand\@biblabel[1]{#1}            
\renewcommand\@makefntext[1]%
{\noindent\makebox[0pt][r]{\@thefnmark\,}#1}
\makeatother 
\renewcommand{\figurename}{\small{Fig.}~}
\sectionfont{\sffamily\Large}
\subsectionfont{\normalsize}
\subsubsectionfont{\bf}
\setstretch{1.125} 
\setlength{\skip\footins}{0.8cm}
\setlength{\footnotesep}{0.25cm}
\setlength{\jot}{10pt}
\titlespacing*{\section}{0pt}{4pt}{4pt}
\titlespacing*{\subsection}{0pt}{15pt}{1pt}

\fancyfoot{}
\fancyfoot[LO,RE]{\vspace{-7.1pt}\includegraphics[height=9pt]{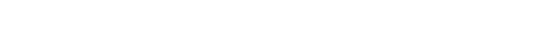}}
\fancyfoot[CO]{\vspace{-7.1pt}\hspace{13.2cm}\includegraphics{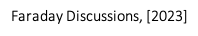}}
\fancyfoot[CE]{\vspace{-7.2pt}\hspace{-14.2cm}\includegraphics{head_foot/RF_new1}}
\fancyfoot[RO]{\footnotesize{\sffamily{1--\pageref{LastPage} ~\textbar  \hspace{2pt}\thepage}}}
\fancyfoot[LE]{\footnotesize{\sffamily{\thepage~\textbar\hspace{3.45cm} 1--\pageref{LastPage}}}}
\fancyhead{}
\renewcommand{\headrulewidth}{0pt} 
\renewcommand{\footrulewidth}{0pt}
\setlength{\arrayrulewidth}{1pt}
\setlength{\columnsep}{6.5mm}
\setlength\bibsep{1pt}

\makeatletter 
\newlength{\figrulesep} 
\setlength{\figrulesep}{0.5\textfloatsep} 

\newcommand{\topfigrule}{\vspace*{-1pt}%
\noindent{\color{cream}\rule[-\figrulesep]{\columnwidth}{1.5pt}} }

\newcommand{\botfigrule}{\vspace*{-2pt}%
\noindent{\color{cream}\rule[\figrulesep]{\columnwidth}{1.5pt}} }

\newcommand{\dblfigrule}{\vspace*{-1pt}%
\noindent{\color{cream}\rule[-\figrulesep]{\textwidth}{1.5pt}} }

\makeatother

\twocolumn[
  \begin{@twocolumnfalse}
{\includegraphics[height=50pt]{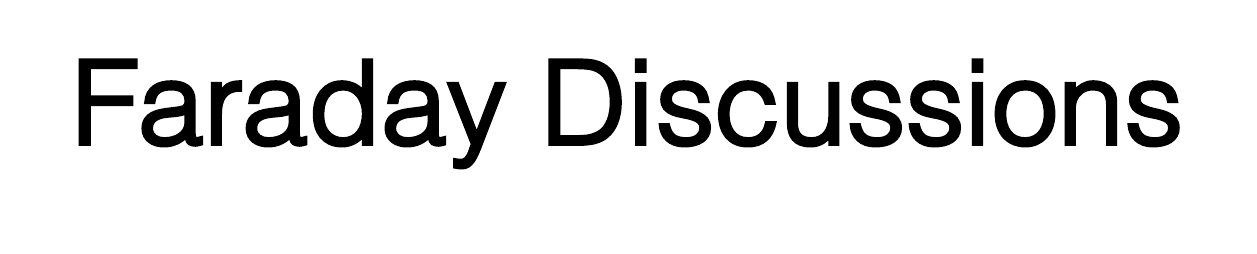}\hfill\raisebox{0pt}[0pt][0pt]{\includegraphics[height=55pt]{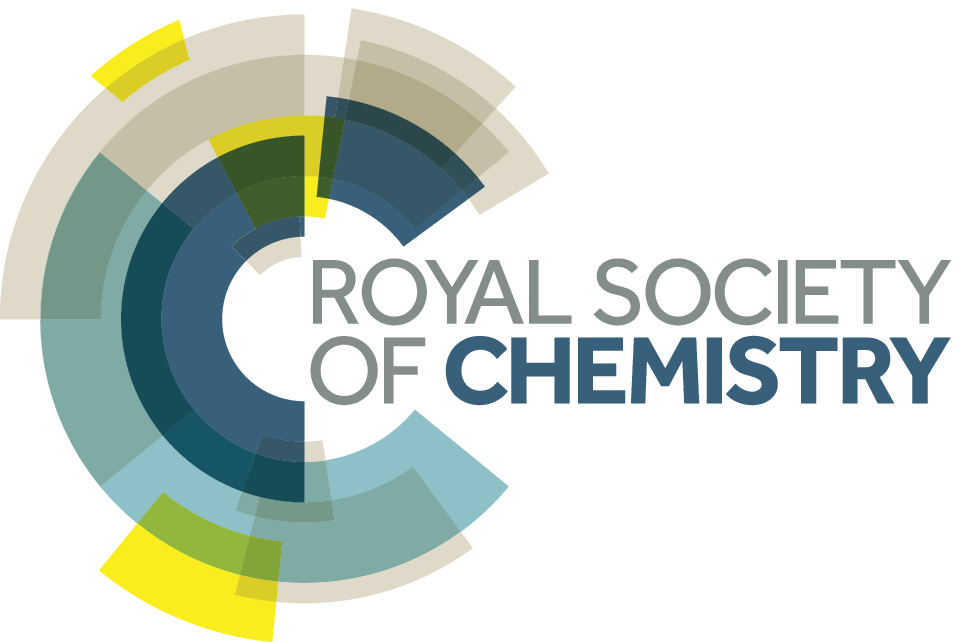}}\\[1ex]
\includegraphics[width=18.5cm]{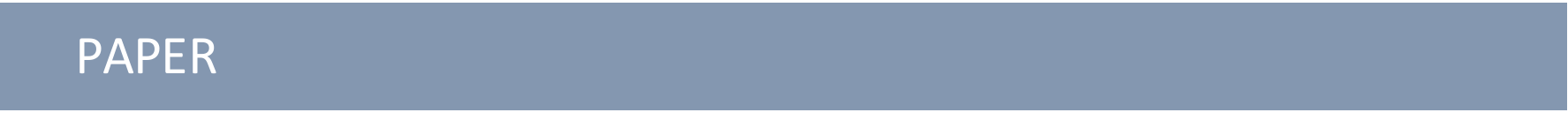}}\par
\vspace{1em}
\sffamily
\begin{tabular}{m{4.5cm} p{13.5cm} p{0.001cm}}

\includegraphics{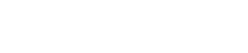} & \noindent\LARGE{\textbf{Chemical Conditions on Hycean Worlds}}
\vspace{0.3cm} & \vspace{0.3cm} \\



 & \noindent\large{Nikku Madhusudhan,\textit{$^{a}$} Julianne I. Moses,\textit{$^{b}$} Frances Rigby\textit{$^{a}$} and Edouard Barrier\textit{$^{a}$}} \\

\includegraphics{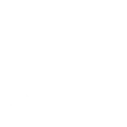} & \noindent\normalsize{
Traditionally, the search for life on exoplanets has been predominantly focused on rocky exoplanets. The recently proposed Hycean worlds have the potential to significantly expand and accelerate the search for life elsewhere. Hycean worlds are a class of habitable sub-Neptunes with planet-wide oceans and H$_2$-rich atmospheres. Their broad range of possible sizes and temperatures lead to a wide habitable zone and high potential for discovery and atmospheric characterization using transit spectroscopy. Over a dozen candidate Hycean planets are already known to be transiting nearby M dwarfs, making them promising targets for atmospheric characterization with the James Webb Space Telescope (JWST). In this work, we investigate possible chemical conditions on a canonical Hycean world, focusing on (a) the present and primordial molecular composition of the atmosphere, and (b) the inventory of bioessential elements for the origin and sustenance of life in the ocean. Based on photochemical and kinetic modeling for a range of conditions, we discuss the possible chemical evolution and observable present-day composition of its atmosphere. In particular, for reduced primordial conditions the early atmospheric evolution passes through a phase that is rich in organic molecules that could provide important feedstock for prebiotic chemistry. We investigate avenues for delivering bioessential metals to the ocean, considering the challenging lack of weathering from a rocky surface and the ocean separated from the rocky core by a thick icy mantle. Based on ocean depths from internal structure modelling and elemental estimates for the early Earth’s oceans, we estimate the requirements for bioessential metals in such a planet. We find that the requirements can be met for plausible assumptions about impact history and atmospheric sedimentation, and supplemented by other steady state sources. We discuss the observational prospects for atmospheric characterisation of Hycean worlds with JWST and future directions of this new paradigm in the search for life on exoplanets.}


\end{tabular}

\end{@twocolumnfalse} \vspace{0.6cm}
 ]

\renewcommand*\rmdefault{bch}\normalfont\upshape
\rmfamily
\section*{}
\vspace{-1cm}



\footnotetext{\textit{$^{a}$~Institute of Astronomy, University of Cambridge, UK. E-mail: nmadhu@ast.cam.ac.uk}}
\footnotetext{\textit{$^{b}$~Space Science Institute, Boulder, CO, USA.}}




\section{Introduction}
\label{sec:intro}
The search for life elsewhere is the holy grail of exoplanet science. While the detection of an atmospheric signature of an Earth-like planet orbiting a sun-like star remains a difficult goal \citep{Arnold2014,Rodler2014,Feng2018}, habitable-zone sub-Neptunes transiting nearby M dwarfs are more accessible; we refer to sub-Neptunes as planet with sizes between Earth and Neptune. Their bright and small host stars lead to larger planet-star size contrasts, making them more conducive to atmospheric characterisation using transit spectroscopy. Traditionally, the search for biosignatures in exoplanets has been focused on rocky exoplanets in the terrestrial habitable zone \citep{Kasting1993,selsis2007,Kopparapu2013}. Like terrestrial planets in the solar system, such planets are expected to have predominantly rocky interior compositions but their atmospheres may be more diverse, ranging from heavy CO$_2$-rich or N$_2$-rich atmospheres to lighter H$_2$-rich atmospheres \citep{Kasting1993,Kopparapu2013,elkins-tanton2008}. Recent transit surveys have discovered a number of super-Earths in the habitable zones of their host stars, notably around nearby M dwarfs, e.g. TRAPPIST-1 \citep{Gillon2017}, LHS~1140 \citep{Dittmann2017,Ment2019,Lillo2020} and TOI-700 \citep{Gilbert2020,Rodriguez2020}. Theoretical studies show that the James Webb Space Telescope (JWST) will have the capability to detect potential atmospheric biosignatures in such planets, albeit with significant investment of observing time \citep{Barstow2016,Snellen2017,Lustig2019}.

 \begin{figure*}
    \centering
    \includegraphics[width=0.9\textwidth]{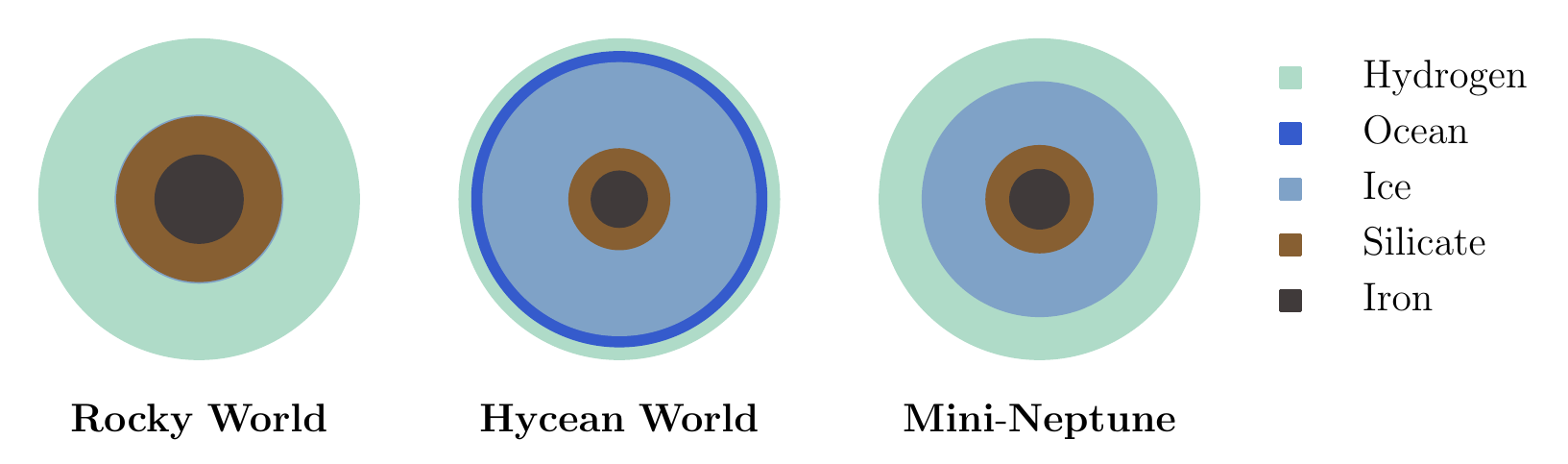}
    \caption{Schematic of degeneracy between internal structures of Hycean worlds with those of other possible planet types with H$_2$-rich atmospheres.}
    \label{fig:subneptunetypes}
\end{figure*}

Recently, a new class of sub-Neptune exoplanets, Hycean worlds, has been proposed to significantly expand and accelerate the search for life elsewhere. Hycean worlds are temperate sub-Neptunes with ocean-covered surfaces underneath H$_2$-rich atmospheres \citep{Madhusudhan2021}. The temperatures and pressures in their oceans allow for habitable conditions similar to those known to sustain life in Earth's oceans. Hycean planets offer significant advantages in the search for life compared to traditionally favoured habitable super-Earths with terrestrial-like atmospheres. Firstly, due to the possibility of a large water fraction such planets can be significantly larger in size for their mass compared to rocky planets, up to 2.6 R$_\oplus$ for 10 M$_\oplus$. Several recent studies have shown temperate sub-Neptunes in this size range to be capable of hosting  surface liquid water and deep oceans for a range of atmospheric and interior conditions\citep{Madhusudhan2020,Piette2020,Nixon2021}. Secondly, the H$_2$-rich atmospheres allow for a much wider habitable-zone for Hycean worlds compared to the terrestrial habitable zone\citep{Madhusudhan2021}. The potential of H$_2$-rich atmospheres to sustain surface habitable temperatures for large orbital separations has also been suggested previously in the context of rocky exoplanets \citep{stevenson1999,pierrehumbert2011}. Hycean planets at large orbital separations are also expected to be conducive for long-term habitability\cite{Mol_Lous2022}. 

Thanks to their physical properties Hycean worlds are more detectable and more accessible to atmospheric characterisation with current facilities compared to rocky habitable planets. The low mean molecular weight  of their hydrogen-rich atmospheres and lower bulk gravity lead to larger atmospheric scale heights for Hycean planets relative to rocky planets of comparable mass. The extended atmospheres along with the large radii make them more favourable for transmission spectroscopy and searching for spectroscopic biomarkers\citep{Madhusudhan2021}. The size range of Hycean planets ($\sim$1--2.6 R$_\oplus$) lie in the sub-Neptune regime ($\sim$1--4 R$_\oplus$), the radius range that dominates the currently known exoplanet population \citep{Fressin2013,fulton2018}. In particular, the Kepler\citep{Borucki2010} and TESS \citep{Ricker2015} missions have discovered tens of temperate sub-Neptunes orbiting nearby M dwarfs, which are excellent targets for transit spectroscopy. Over a dozen such planets have already been identified as candidate Hycean worlds with high potential for atmospheric follow-up \citep{Madhusudhan2021,Fukui2022,Kawauchi2022,Evans2023,Piaulet2023}. Simulation studies with JWST show that such planets are indeed excellent candidates for detections of key volatile species and potential biomarkers in their atmospheres \citep{Madhusudhan2021,Phillips2021,Phillips2022,Leung2022}.

Characterising the atmospheric composition is essential to identify a Hycean world. A degenerate set of internal structures can generally explain the observed mass, radius and equilibrium temperature of a candidate Hycean world \citep{Madhusudhan2020}. As shown in Fig.~\ref{fig:subneptunetypes}, these solutions include rocky worlds with large H$_2$-rich envelopes or mini-Neptunes which contain both a rocky layer and an icy mantle along with a significant H$_2$-rich envelope, similar to the ice giants Uranus and Neptune in the solar system. Neither of these two scenarios would support a habitable surface, given the high temperature and pressure expected underneath the thick H$_2$ envelope. Hycean worlds would lie between these two extremes, with a thin H$_2$-rich atmosphere that allows for a liquid water layer underneath at habitable temperatures and pressures \citep{Madhusudhan2021}. Recent studies have suggested using atmospheric compositions to infer the presence of surfaces in sub-Neptunes with H$_2$-rich atmospheres\citep{Yu2021,Hu2021,Tsai2021}. Thus, resolving the degeneracies in internal structure and reliably identifying a Hycean world would require an understanding of the expected atmospheric chemical processes and the observable compositions. 

While Hycean worlds offer the required thermodynamic conditions for oceanic life their possible chemical conditions have not been explored in detail. A central question in determining the possibility of life on a planet is whether its environment has the potential for the origin and sustenance of life. The minimum ingredients required for life on Earth are known to be\citep{McKay2014,Cockell2016}: (a) an energy source, (b) liquid water, and (c) various bioessential elements. Hycean worlds clearly satisfy the requirements for the presence of an energy source and liquid water  with the presence of the host star and an abundant source of water. However, as is common to all ocean worlds, including icy moons in the solar system\cite{Hendrix2019}, the availability of chemical inventory accessible for seeding and sustaining life is a key challenge \cite{Lammer2009, Noack2016, Lingam2018,Journaux2020}. In particular, Hycean worlds, with their planet-wide oceans, hinder access to bioessential elements resulting from geochemical cycles involving weathering of rocky surfaces that would be natural for a rocky planet. The large water fraction possible on such planets, as for other ocean worlds and sub-Neptunes in general, can result in a thick layer of high-pressure ice disconnecting the ocean from the rocky core and limiting access to bioessential nutrients\citep{Lammer2009,Maruyama2013,Noack2016,Seager2021}. At the same time, other possible sources of nutrient delivery remain possible, including delivery through impacts \citep{Seager2021,Cockell2016}, extraterrestrial dust \citep{Seager2021,Cockell2016} and potential transport of nutrients from the inner core through a permeable ice layer \citep{Choblet2017, Kalousova2018, KS2018,Hernandez2022,Lebec2023}. 
In addition to bioessential elements, the availability of pre-biotic organic molecules also plays an essential role in the origin of life. Pre-biotic molecules such as HCN are thought to have been critical for abiogenesis in the Early Earth \citep{Rimmer2018}. Whether such molecules can be delivered through extraterrestrial impactors or need to be created in situ, or both, remains an open debate \citep{Parkos2018}.

In the present work, we explore the chemical conditions possible on Hycean worlds. Using a canonical model of a Hycean world, we investigate the possible molecular inventory resulting from photochemical and kinetic processes in the atmosphere. We also explore sources of bioessential elements possible on such planets. In what follows, we describe our methods in section~\ref{sec:methods}, including the atmospheric structure, photochemistry, and internal structure models. We present our results and discuss the chemical conditions possible for life on Hycean worlds in section~\ref{sec:results}. We discuss the observational prospects for identifying and characterising Hycean atmospheres with JWST in section~\ref{sec:jwst}. We summarise our findings and discuss future directions in section~\ref{sec:summary}. 

 \begin{figure*}
    \centering
    \includegraphics[width=0.4\textwidth]{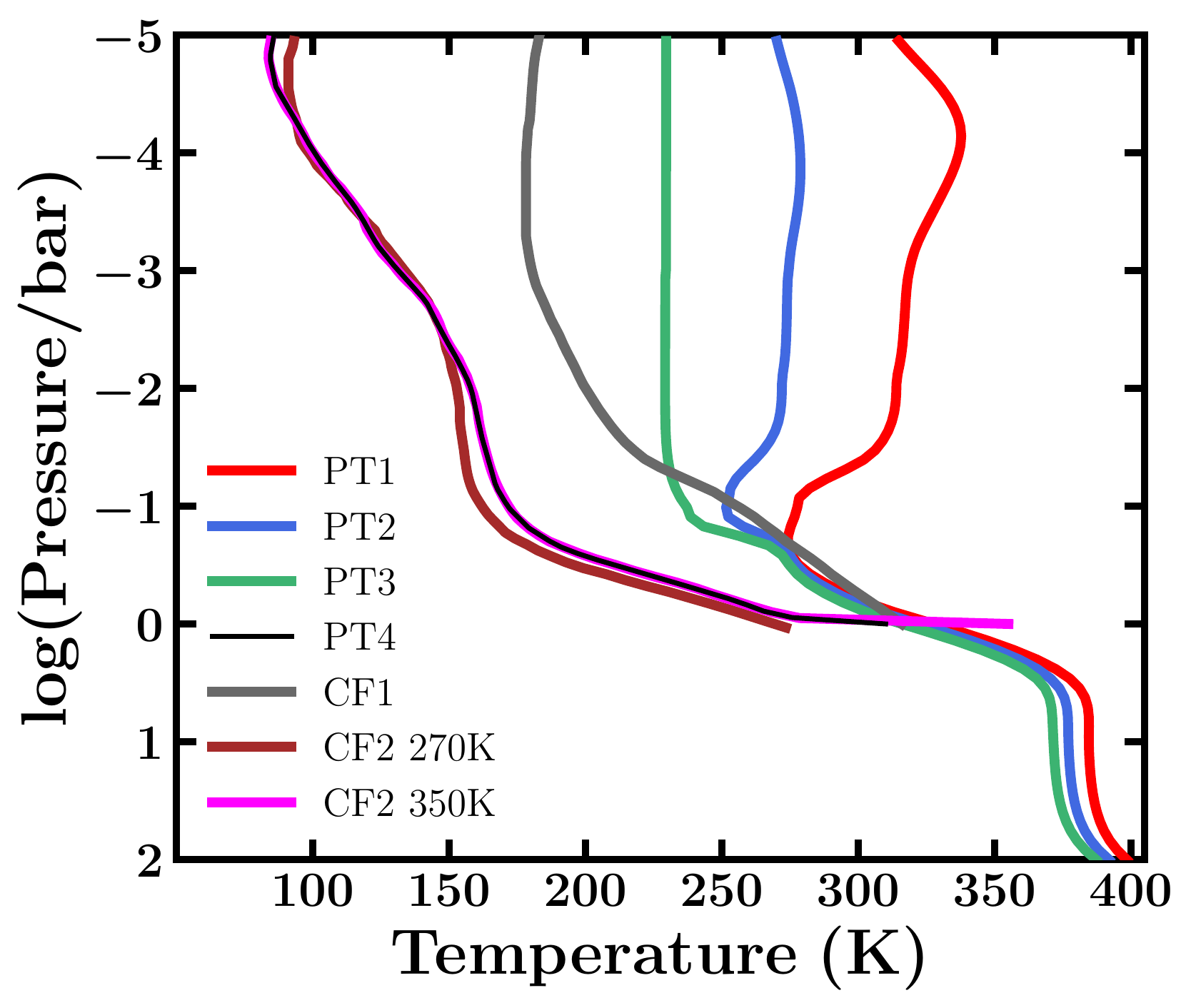}
    \includegraphics[width=0.49\textwidth]{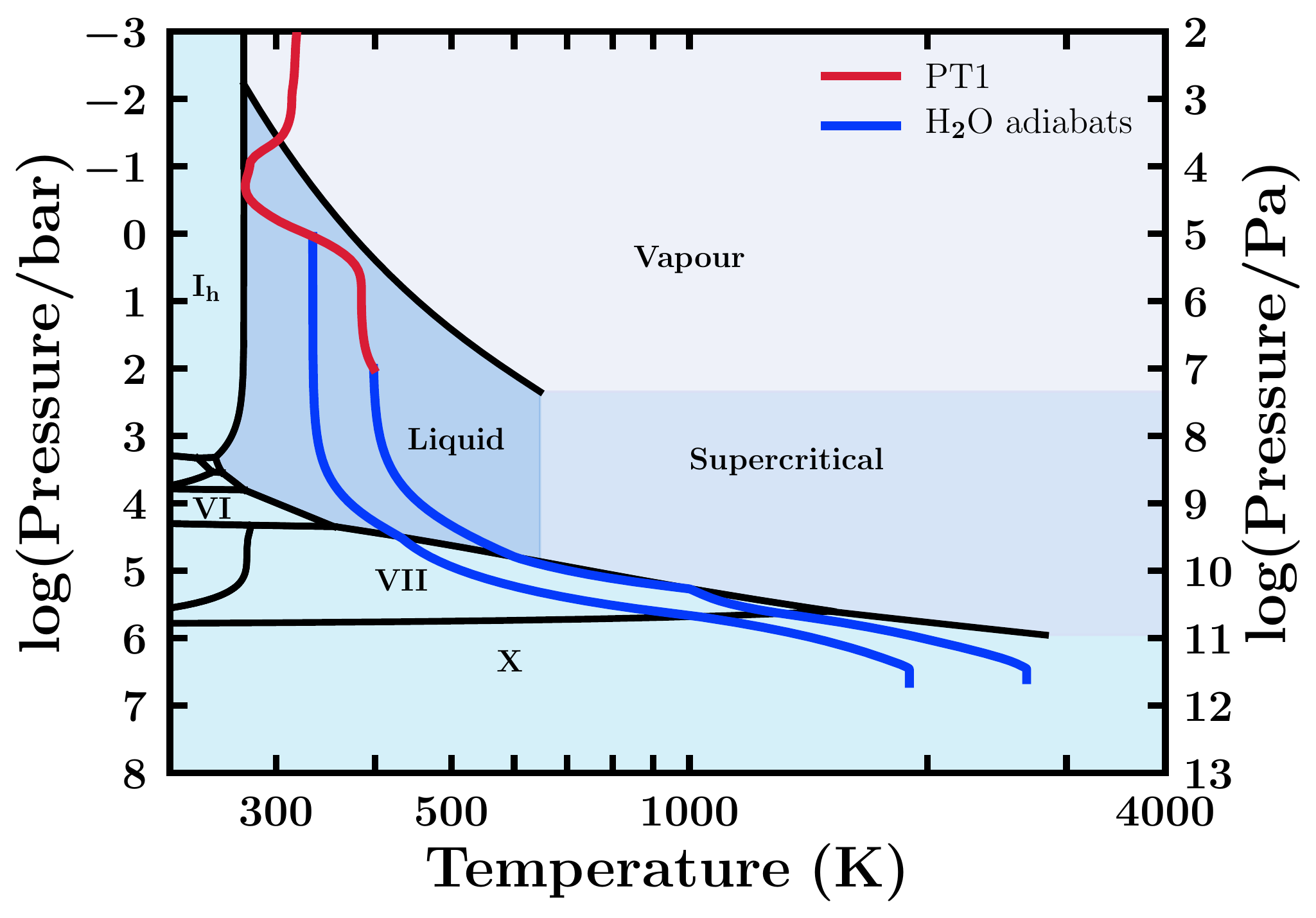}
    \caption{Left: Model Pressure-Temperature ($P$-$T$) profiles of a canonical Hycean world considered in this study (see section~\ref{sec:results_structure}). The profiles PT1-PT3 consider scattering due to hazes modelled as enhanced Rayleigh scattering. CF1 and CF2 refer to $P$-$T$ profiles from previous studies\cite{Hu2021,Innes2023} assuming cloud/haze free (CF) model atmospheres for different conditions as discussed in section~\ref{sec:results_structure}. Two profiles are shown for CF2\cite{Innes2023} with different surface temperatures, and we consider a profile PT4 which has an intermediate surface temperature between the two CF2 profiles. Right: Internal temperature structure and the H$_2$O phase diagram. The black curves denote the different phase boundaries. The blue curves show H$_2$O adiabats starting at ocean surface pressures of 1 bar and 100 bar. The red curve shows the PT1 temperature profile of the canonical atmosphere from the left panel for reference.}
    \label{fig:pt_profiles}
\end{figure*}

\section{Methods} 
\label{sec:methods}

We explore the possible chemical conditions in a canonical Hycean planet focusing on the molecular composition of its atmosphere and the availability of bioessential elements in its ocean. We model the atmospheric pressure-temperature structure using a self-consistent radiative-convective equilibrium model, the atmospheric composition using a photochemical model, and the ocean properties using an internal structure model. Here we describe the model considerations. 

\subsection{Atmospheric Modelling}
\label{sec:methods_atmos}

We consider a plane-parallel H$_2$-rich atmosphere in radiative-convective equilibrium under Hycean conditions. We model the atmosphere using the GENESIS self-consistent modelling framework \citep{Gandhi2017} adapted for sub-Neptune atmospheres \citep{Madhusudhan2020,Piette2020,Madhusudhan2021}. This model self-consistently computes the temperature structure in radiative-convective equilibrium using the Rybicki scheme considering external irradiation, internal flux, possible scattering due to clouds/hazes, opacities from different chemical compositions (in chemical equilibrium or for a given composition), and day-night energy transport. The model computes line-by-line radiative transfer using the Feautrier method \citep{Hubeny2017} and the Discontinuous Finite Element method \citep{Castor1992}. For given system parameters and chemical composition, the model self-consistently computes the temperature profile, flux profile, and the emergent spectrum of a planet. 

We assume a canonical Hycean planet with the bulk properties of the habitable-zone sub-Neptune K2-18~b, which has been suggested as a candidate Hycean planet \citep{Madhusudhan2020,Madhusudhan2021}. We adopt the system parameters reported by Benneke et al \citep{Benneke2019}, with a planet mass of 8.63 M$_\oplus$ \citep{Cloutier2019} and radius of 2.61 R$_\oplus$ \citep{Benneke2019} orbiting an M3V host star. The thermodynamic conditions on Hycean worlds can span a large range, with the ocean surface temperatures of 273-400 K and surface pressures of 1-10$^3$ bar. For our canonical model, we consider a limiting case of a warm Hycean world in which the ocean surface is at a temperature ($T$) of $\sim$400 K and a pressure ($P$) of 100 bar. This choice is governed by the fact that most of the Hycean candidates currently identified are at zero-albedo equilibrium temperatures around 400~K. The bottom of the atmosphere is set at 100 bar, denoting the pressure at the ocean-atmosphere boundary, and the top at 10$^{-6}$ bar. We derive the temperature profile with the required temperature and pressure at the bottom boundary by a combination of model parameters, which involve the incident irradiation, internal temperature, and scattering; an exploration of this parameter space is discussed in recent studies\citep{Madhusudhan2020,Piette2020,Madhusudhan2021}. 

We first consider our canonical planet to be irradiated with the maximal flux received by K2-18~b, i.e at the sub-stellar point with no day-night energy redistribution ($f_{\rm r}$), which corresponds to an irradiation temperature of 331 K. Alternately, assuming a nominal $f_{\rm r}$ of 0.35 this flux would correspond to an equivalent irradiation temperature of 369 K. We assume a nominal internal temperature of 30 K. The primary source of molecular opacity in this canonical model atmosphere is due to H$_2$O \citep{barber2006,rothman_hitemp_2010}. We also include H$_2$-H$_2$ and H$_2$-He collision-induced absorption\citep{borysow1988,orton2007,abel2011,richard_new_2012}. We assume the H$_2$O abundance to be saturated in the lower atmosphere until a pressure of 0.1 bar at which point the H$_2$O abundance is assumed to be quenched higher up. We explore atmospheric pressure-temperature ($P$-$T$) profiles obtained using different levels of scattering in the atmosphere. Various sources of scattering can influence the $P$-$T$ profile, including clouds and hazes of different compositions \citep{Madhusudhan2020,Piette2020,Madhusudhan2021}. Here, we follow the approach of recent studies\citep{Piette2020,Madhusudhan2021} and consider extinction due to hazes modelled as an enhanced Rayleigh scattering, with H$_2$ Rayleigh scattering enhanced by a multiplicative factor which is a free parameter in the model. We also consider $P$-$T$ profiles from other recent studies\citep{Hu2021,Innes2023} that assumed cloud/haze-free atmospheres for Hycean worlds. We discuss the specific $P$-$T$ profiles used in this work in section~\ref{sec:results_structure}.

\subsection{Chemical Kinetics Modeling}
\label{sec:methods_photochem}
To investigate the possible atmospheric composition of our canonical Hycean world, we use the Caltech/JPL 1D KINETICS photochemical model \cite{allen81,yung84,moses11} to solve the coupled continuity equations for the chemical production, loss, and vertical transport of atmospheric constituents. First, we consider models with a zero-flux boundary condition at the bottom of the atmosphere to investigate how the composition evolves with time  chemically, when the atmospheric temperature at the ocean surface is too low to allow full chemical recycling of the photochemical products back into their original equilibrium constituents.  Similar to previous investigations \citep{Yu2021,Tsai2021}, these toy models start with a thermochemical equilibrium composition for different assumptions about bulk elemental abundances, and then the composition is allowed to evolve with time under stellar UV irradiation, assuming zero flux boundary conditions for all species (i.e., no atmospheric escape or external input at the top of the atmosphere, and no chemical exchange with or outgassing from the surface). The zero-flux assumption allows for a quick look at the timing and behavior of the transition from the assumed reduced initial atmosphere accreted from the protoplanetary disk --- with H$_2$O, CH$_4$, and NH$_3$ containing the bulk of the oxygen, carbon, and nitrogen in this cool, H$_2$-rich atmosphere --- to the CO$_2$, CO, H$_2$O, N$_2$ dominant end state (also still within a H$_2$-rich atmosphere) that occurs when the more photochemically fragile reduced forms of the elements are converted to  molecules with stronger chemical bonds \cite{Yu2021,Tsai2021}. Of course, chemical exchange between the surface and atmosphere will occur on a real planet, with outgassing of dissolved CO$_2$ being particularly important for ocean worlds \citep{Hu2021,Kite2018}, which we also consider as discussed below. Table~\ref{tab:chem_models} shows the cases reported in this work, with the zero-flux models corresponding to Cases 1-6. 

These zero-flux models use the fully reversed reaction mechanism of \citet{moses13} that includes 92 species containing the elements C, N, O, and H that interact with each other via $\sim$1650 reactions; further model details are provided in \citet{moses13} and \citet{Yu2021}.  As is discussed by \citet{Hu2021}, sulfur species are expected to be largely sequestered in the ocean on such planets, and volcanic outgassing is expected to be shut down by the overlying high pressures at the rock-ocean boundary \citep{Kite2009,Kite2018}. The eddy diffusion coefficient $K_{zz}$ in our model atmospheres is assumed to vary with pressure $P$ according to $K_{zz}$ = $5.6 \, \times \, 10^{4} P^{-0.5}$ cm$^2$ s$^{-1}$, with $P$ in bar (\citep[see][]{moses22tremgrid}), with $K_{zz}$ capped at 10$^{10}$ cm$^2$ s$^{-1}$ in the upper atmosphere, and $K_{zz}$ = 10$^6$ cm$^2$ s$^{-1}$ at $P$ $>$ 0.5 bar.  The stellar flux is taken from the HAZMAT database \cite{peacock20}, and we investigate results assuming both a young (45 Myr) and old (5 Gyr) M dwarf of 0.45 solar mass, with a medium EUV level for that age/mass; the younger star has a higher EUV flux.  One change from \citet{Yu2021} is that we consider H$_2$O condensation in these models, following procedures outlined in \citet{moses00b}.  Water vapor is saturated at the bottom (ocean) boundary. With some of the thermal profiles, H$_2$O will condense again higher up in the atmosphere if the H$_2$O vapor mixing ratio exceeds local saturation.  Condensation/evaporation are assumed to be zero above the minimum in the saturation vapor mixing ratio near the tropopause (if one exists), such that the water vapor at higher altitudes never exceeds that minimum mixing ratio; i.e., we assume condensed water droplets or ice particles are not carried to higher altitudes where they might re-evaporate.  Model results are presented at various points in time as the composition evolves.

We use these zero-flux models to gain insight into the potential atmospheric composition at various stages in the planet’s history.  All the models pass initially through an organic-rich phase, whose lifetime depends on the atmospheric thermal structure, the pressure at the ocean surface, and the UV flux from the star.  All models end up with at least partial conversion of the original CH$_4$ and NH$_3$ to CO$_2$, CO, N$_2$, and heavier organic molecules.  We use the output from these simple models to define boundary conditions for more realistic photochemical models designed to explore the atmospheric composition of a Hycean planet with a surface ocean.  

The atmospheric composition of these more realistic Hycean-world models will depend strongly on boundary conditions, which are not known \textit{a priori} for these planets.  Guided by our zero-flux model results and other discussions in the literature \cite{Kasting1990,Arney2016,Ranjan2020,Yu2021,Tsai2021,Hu2021}, we examine an early organic-rich scenario where the atmosphere is assumed to start with a Neptune-like 100$\times$ solar metallicity composition in thermochemical equilibrium, with H$_2$O controlled by saturation at the bottom of the model.  Mixing ratios of H$_2$O, NH$_3$, CH$_4$, and He are fixed at the lower boundary at these equilibrium (or saturated for H$_2$O) abundances. Fixing the mixing ratio in these models in combination with photochemical loss results in an upward flux of O, C, N that in steady state must end up being balanced by a loss of photochemical products through the oceanic lower boundary.  Zero flux is assumed for all species at the top of the atmosphere.  Based on several early Earth models and/or ocean-world models \cite{Kasting1990,Arney2016,Ranjan2020,Hu2021}, we have chosen deposition velocities $v_{d}$ at the lower boundary of zero for O$_2$, N$_2$, 
and several long-lived C$_2$H$_x$--C$_4$H$_x$ hydrocarbons; $v_{d}$ = 10$^{-8}$ cm s$^{-1}$ for CO; $v_{d}$ = 10$^{-5}$ cm s$^{-1}$ for CO$_2$, C$_2$H$_6$, C$_3$H$_8$, C$_4$H$_{10}$, and stable hydrocarbons with 5 or more carbon atoms (precursors to organic hazes); $v_{d}$ = 10$^{-3}$ cm s$^{-1}$ for H$_2$CO, CH$_3$OH, H$_2$CCO, CH$_3$CHO, HCN, CH$_3$CN, HC$_3$N, CH$_3$NH$_2$, NO, HNCO; all other species are assumed to flow through the boundary at the maximum possible rate, given by $K_{zz}$ divided by the atmospheric scale height.  For our canonical Hycean world, the scale height at the surface is larger than that on Earth, and the maximum $v_d$ is $\sim$0.22 cm s$^{-1}$.  We use the stellar flux from the younger star for this organic-rich ``early-era’’ model. This model is used for Cases 7 and 8 in Table~\ref{tab:chem_models}.

We also consider ``late-era'' scenarios where the reduced parent species have already been depleted through conversion to more stable photochemical products, such that the mixing ratios of CO$_2$, N$_2$, H$_2$O, and He are fixed at the lower boundary. These cases are similar to the thin-atmosphere ocean-world models presented by \citet{Hu2021}, where dissolved CO$_2$ is expected to be the dominant carbon phase in the ocean\citep{Kite2018}, and the amount of atmospheric CO$_2$ is controlled by its equilibrium with the ocean, which in turn depends on the oceanic pH. \citet{Hu2021} explored a range of possible fixed CO$_2$ mixing ratios in their model, and we also consider three cases: (1) a 100-bar, low-CO$_2$ case with a CO$_2$ partial pressure of 10$^{-4}$ bar (i.e., a mixing ratio of 10$^{-6}$ for our 100-bar surface pressure), within the range of the lower bound calculated for a pH of 9--10 by \citet{Hu2021}, (2) a 100-bar, intermediate CO$_2$ mixing ratio (along with N$_2$) that matches the mixing ratios at the lower boundary of our zero-flux model runs at the 5 Gyr time period (i.e. 7.2$ \, \times \, 10^{-3}$ for N$_2$ and 1.37$ \, \times \, 10^{-2}$ for CO$_2$ in our warmest model), and (3) a 1-bar, high-CO$_2$ case (10\% CO$_2$, 1\% N$_2$) that aligns with the high-CO$_2$ case from \citet{Hu2021}.  The H$_2$O mixing ratios in these models are fixed at saturation values at the lower boundary.  The other boundary conditions are the same as in the organic-rich early-era model, except that CH$_4$ is assumed to have $v_d$ = 0 and NH$_3$ diffuses through the bottom boundary at the maximum possible rate.  These ``late era'' models use the stellar flux from the older M dwarf star and are run to steady state. These models are represented by Cases 9-11 in Table~\ref{tab:chem_models}. We discuss our results in section~\ref{sec:results}.

\subsection{Internal Structure Modelling}
The internal structure is modelled following the approach used for Hycean candidates in recent  studies \citep{Madhusudhan2012,Madhusudhan2020,Nixon2021}. For a given planet mass ($M_{\mathrm{p}}$) and the mass fractions of the constituent layers ($x_i$), the planet radius ($R_{\mathrm{p}}$) is obtained by solving the structure equations, i.e. of mass continuity and hydrostatic equilibrium. We assume four-layers -- an H$_2$/He-rich atmosphere, an H$_2$O layer, a silicate outer core and an Fe inner core. For each layer, the equation of state (EOS) and temperature profile are specified as described below. The structure equations, the adopted EOSs and the $P$-$T$ profiles form a set of coupled differential equations which are solved via a fourth-order Runge-Kutta method.

In the H$_2$-rich atmosphere the pressures and temperatures required to produce habitable surface conditions are adequately low that the ideal gas EOS is sufficiently accurate. More generally, for the H/He layer we use the EOS from \citet{Chabrier2019} given the atmospheric $P$-$T$ profile. The surface is taken to be at the ocean-atmosphere boundary, or H$_2$O-H$_2$/He boundary (HHB)\citep{Madhusudhan2020}. In the H$_2$O layer, we adopt an adiabatic $P$-$T$ profile, with the initial pressure and temperature set by the HHB. The temperature-dependent EOS for H$_2$O is compiled from multiple sources -- see previous studies \citep{Thomas2016,Nixon2021} for a full description of these. The EOSs we adopt for the core \citep{Seager2007} are in the form of a Birch-Murnaghan EOS \citep{Birch1952} for MgSiO$_3$ perovskite \citep{Karki2000} and Fe \citep{Ahrens2000}, which are temperature-independent. The core is assumed to be Earth-like in composition, with an Fe mass fraction of 0.33. 

\begin{table*}
\centering
\begin{tabular}{|c|c|c|c|c|c|c|c|c|c|}
\hline 
Case  & $P-T$ $\rm {profile}$  & $P_{\rm b}$/bar &  Star &  Boundary Condition & H$_2$O & CH$_4$ & NH$_3$ & CO$_2$ & CO \\
\hline
1  &  PT1  &      100 &     old &       zero flux & 2.4E-02 & 1.1E-02 & 4.1E-05 & 1.3E-02 &	1.6E-05\\ 
2  &  PT2  &      100 &     old &       zero flux & 1.0E-02 & 6.6E-03 & 1.5E-04 & 4.5E-03 &	5.6E-05\\ 
3 &  PT3  &       100 &     old &       zero flux & 2.8E-03 & 2.5E-03 & 3.2E-04 & 3.5E-04 &	2.5E-04\\ 
4 &  PT1  &         1 &     old &       zero flux & 2.4E-02 & 6.8E-06 & 3.0E-09 & 4.1E-02 &	1.9E-03\\ 
5 &  PT3  &         1 &     old &       zero flux & 2.3E-03 & 8.2E-05 &	3.9E-13	& 4.2E-02 &	9.3E-04\\ 
6 &  PT4  &         1 &     old &       zero flux & 3.1E-10 & 7.8E-05 & 2.2E-07 & 9.4E-15 & 3.0E-09\\
\hline
7* &  PT1  &  100 &     young &      fixed H$_2$O, CH$_4$, NH$_3$ & 2.5E-02 &	5.0E-02	&	1.5E-02	&	8.5E-05 &	7.9E-05 \\
8*    &  PT4  &    1 &     young &       fixed H$_2$O, CH$_4$, NH$_3$ & 3.0E-10 & 4.9E-02 & 4.7E-03 & 1.8E-19 & 2.4E-10 \\
9    &  PT1  &       100 &     old &       fixed N$_2$, H$_2$O; CO$_2$=1.4E-02  & 2.4E-02 &	1.2E-04	&	1.7E-10	&	1.3E-02 &	4.9E-06 \\ 
10    &  PT1  &     100 &     old &       fixed N$_2$, H$_2$O; CO$_2$ = 1.0E-06 & 2.4E-02 &	3.5E-06	&	2.8E-10	&	1.0E-06 &	2.5E-08 \\
11    &  PT4  &    1 &     old &       fixed N$_2$, H$_2$O; CO$_2$ = 1.0E-01 & 3.1E-10	& 5.5E-08 & 6.3E-13 & 1.0E-01 &	6.4E-03 \\	
\hline 
\end{tabular}
\caption{Model assumptions and volume mixing ratios of prominent molecules at 1 mbar for the Hycean cases explored with the photochemical calculations in this work (see section~\ref{sec:methods_photochem}). The initial metallicity is assumed to be 100$\times$ solar. $P_{\rm b}$ is the pressure at the lower boundary of the atmosphere. The boundary conditions are those considered at the lower boundary. The first six cases consider zero-flux boundary conditions, i.e. an unreactive surface, whereas the last five cases consider boundary conditions as discussed in section~\ref{sec:methods_photochem}. Cases 7* and 8* compute chemistry at an age of 28 Myr in the early evolution of the planet, assuming reduced atmospheric boundary conditions representative of 100$\times$ solar metallicity. The remaining cases are evolved until 5 Gyr. The last five columns show the volume mixing ratios at 1 mbar of prominent CNO molecules (H$_2$O, CH$_4$, NH$_3$, CO$_2$ and CO) expected in H$_2$-rich atmospheres. The H$_2$O abundance is controlled by saturation at the lower boundary, that is mixing ratios of $\sim$2$\times$10$^{-2}$ for the 100-bar cases, except when condensation occurs at higher altitudes (e.g., PT2, PT3 and PT4 profiles), where the saturation minimum becomes $\sim$1$\times$10$^{-2}$ for PT2, $\sim$3$\times$10$^{-3}$ for PT3, and lower for PT4. We note that cooler temperature profiles can decrease the H$_2$O abundance further and affect the abundances of other molecules, especially decrease the CO$_2$ abundance.}
\label{tab:chem_models}
\end{table*}

\section{Results}
\label{sec:results}

The potential for life on a habitable-zone planet depends on a wide range of properties, including the thermodynamic and chemical conditions in both the atmosphere as well as the interior. It has been suggested previously that Hycean planets allow for a wide range of thermodynamic conditions in their atmospheres and oceans, similar to conditions known to be conducive for life in Earth’s oceans \citep{Madhusudhan2021}. Here we explore the possible chemical composition of a canonical Hycean atmosphere and the availability of bioessential elements in its ocean.

 \begin{figure*}[ht]
    \centering
    \includegraphics[width=0.48\textwidth]{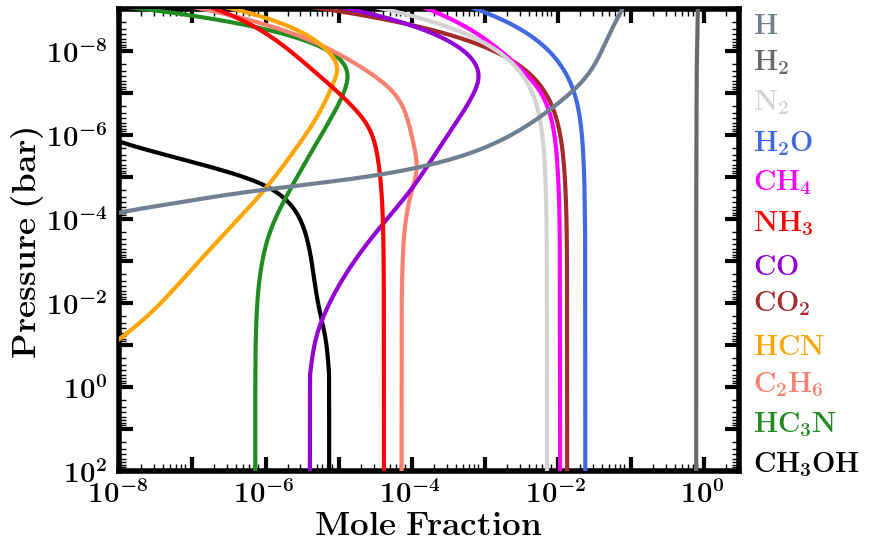}
    \includegraphics[width=0.48\textwidth]{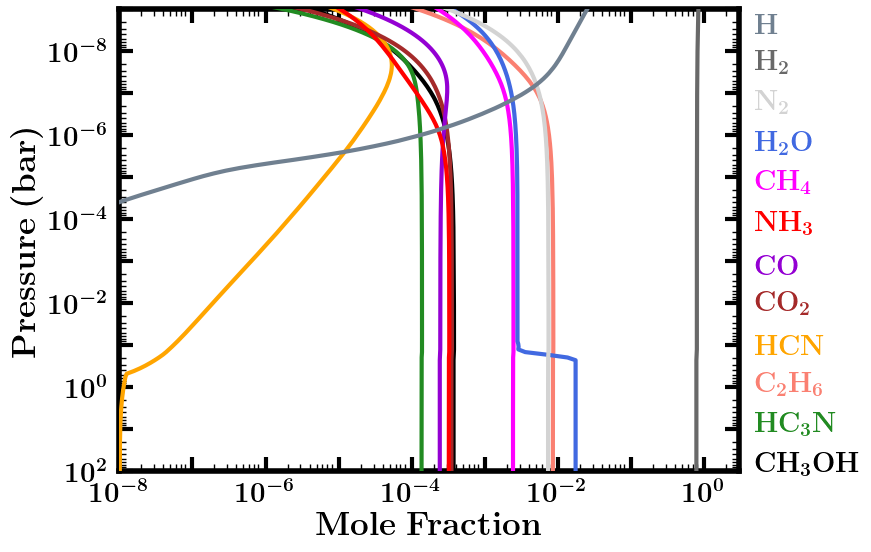}
    \includegraphics[width=0.48\textwidth]{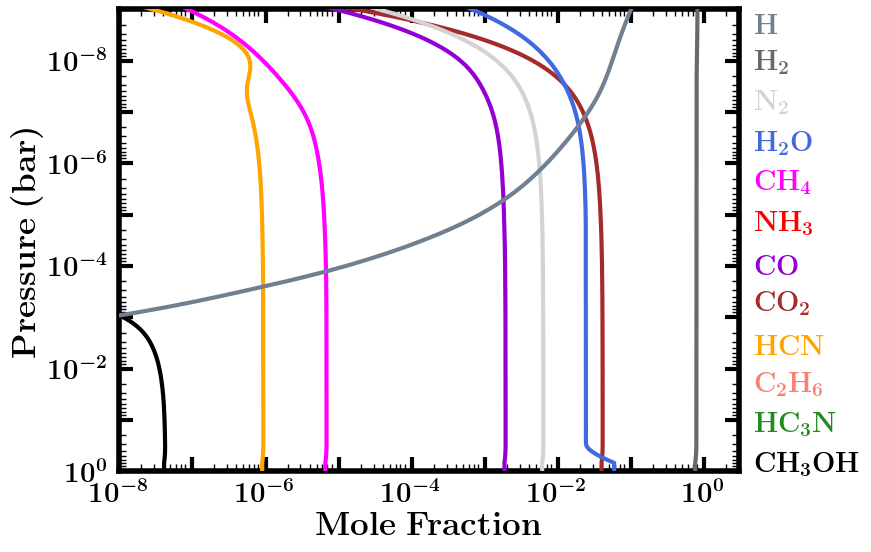}
    \includegraphics[width=0.48\textwidth]{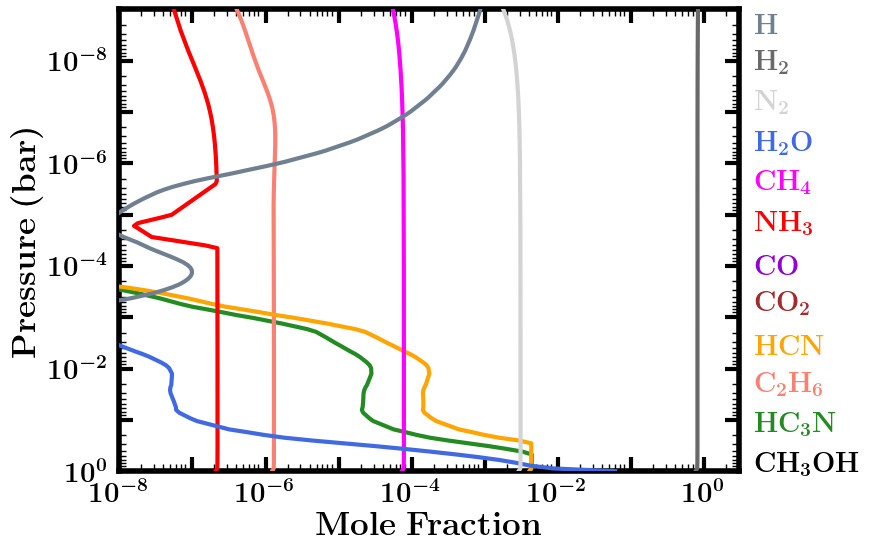}
    \caption{Fully evolved (steady-state) volume mixing-ratio profiles of prominent volatile species with zero-flux boundary conditions (inactive surface) for a 100$\times$ solar composition atmosphere and surface pressure of (Top) 100 bar and (Bottom) 1 bar: (Top Left) for the PT1 temperature profile in Figure~\ref{fig:pt_profiles}, corresponding to Case 1 in Table~\ref{tab:chem_models}; (Top Right) for the PT3 temperature profile, corresponding to Case 3; (Bottom Left) for  the PT1 temperature profile, corresponding to Case 4; (Bottom Right) for the PT4 temperature profile, corresponding to Case 6. Note that the sharp decline in H$_2$O, HCN, HC$_3$N, and NH$_3$ with altitude in Case 6 is due to condensation.}
    \label{fig:pt1_3_sidebyside}
\end{figure*}

\subsection{Atmospheric and Interior Structure}
\label{sec:results_structure}
We consider a canonical Hycean planet and investigate the possible chemical inventory in its atmosphere for a range of physical conditions. The temperature structure of the atmosphere is arguably the central factor influencing the habitability at the ocean surface. A wide range of atmospheric temperature structures are possible in temperate sub-Neptunes, including Hycean worlds, depending on the stellar irradiation, internal temperature, atmospheric composition, and scattering due to clouds/hazes \citep{Madhusudhan2020,Piette2020,Scheucher2020,Blain2021,Charnay2021,Madhusudhan2021,Pierrehumbert2023,Innes2023}. In particular, an adequate albedo due to clouds/hazes is required to maintain a liquid water surface for most candidate Hycean worlds\citep{Madhusudhan2020,Piette2020,Madhusudhan2021}. An atmosphere without strong scattering due to clouds/hazes would generally render the surface too hot to maintain a habitable ocean \cite[e.g.][]{Piette2020,Scheucher2020,Blain2021,Pierrehumbert2023,Innes2023}. The model considerations in this work are discussed in section~\ref{sec:methods_atmos}, based on the bulk properties of the candidate Hycean planet K2-18~b. We consider four $P$-$T$ profiles for our models.  First, we construct three profiles (PT1, PT2, and PT3) shown in Fig.~\ref{fig:pt_profiles}, considering  scattering due to hazes modelled with Rayleigh enhancement factors of 900, 1100 and 1300, respectively. These profiles represent a range of possible atmospheric temperatures but all have a maximum temperature of $\sim$400 K and pressure of 100 bar at the ocean surface. As discussed in section~\ref{sec:methods_atmos}, our choice of this limiting case is driven by expectations for most of the candidate Hycean worlds currently known. This surface temperature is also close to the maximum ocean temperature of 395 K at which life is known to survive in Earth's oceans \cite{Rothschild2001,Merino2019}. Any cooler Hycean worlds would be even more conducive to habitability, making the present case a conservative choice. We explore additional cases in which the PT1 and PT3 profiles are limited to the 1 bar pressure level, as shown in Table~\ref{tab:chem_models}, representing shallower atmospheres with the ocean surface at 1 bar and temperatures of $\sim$310-330 K.

We also consider a $P$-$T$ profile based on a cloud/haze-free (CF) model atmosphere of K2-18~b as pursued in recent works\citep{Hu2021,Innes2023}. \citet{Hu2021} consider an H$_2$-rich atmosphere of K2-18~b with Earth-like equivalent insolation, achieved assuming a Bond albedo of 0.3, but otherwise considering no clouds/hazes in the model. \citet{Innes2023} report CF model $P$-$T$ profiles for a K2-18~b-like planet with surface temperatures of 270 K and 350 K at 1 bar pressure, as shown in Fig.~\ref{fig:pt_profiles} and denoted by ``CF2 270 K'' and ``CF2 350 K'', respectively. These correspond to the inner edge of the CF Hycean habitable zone around an M dwarf similar to K2-18, which for a 1-bar atmosphere is at 0.28 au\cite{Innes2023}, i.e. corresponding to the runaway greenhouse limit for a steam-dominated lower atmosphere. We note that for these profiles a significant part of the atmosphere lies below freezing temperatures, making the presence of clouds/hazes and, hence, a high albedo arguably inevitable and inherently inconsistent with the assumption of a CF atmosphere. Nevertheless, it is still instructive to investigate the effect of photochemistry in such cool Hycean atmospheres. We consider a variant of the CF2 350 K profile, denoted as PT4, restricting the surface temperature to 310 K, which is intermediate between the two CF2 profiles, avoids steam domination, and is close to the other profiles shown in Fig.\ref{fig:pt_profiles} at 1 bar, enabling a comparative study. Overall, we consider four $P$-$T$ profiles (PT1-PT4) for our photochemical models as shown in Table~\ref{tab:chem_models}.

We also use the atmospheric temperature profiles to define the outer boundary condition for the internal structure model as pursued previously for K2-18~b\citep{Madhusudhan2020}. The internal structure model is described in section \ref{sec:methods}. The goal of the internal structure modelling is to estimate the range of ocean depths possible in our canonical Hycean world, which is then used to estimate the inventory of bioessential elements required in the ocean. For this purpose, we use the hottest of the four PT profiles (PT1), which leads to somewhat deeper oceans\citep{Nixon2021} and therefore requires a higher nutrient budget, i.e. our conservative case. We consider two different surface pressures for the ocean and estimate the range of ocean depths possible given the 1$\sigma$ bounds on the mass and radius of the planet.  For a 1 bar surface, we obtain a minimum ocean depth of 120 km, for an upper bound mass for K2-18~b of 9.98 M$_{\oplus}$ \citep{Cloutier2019,Benneke2019}. From the atmospheric $P$-$T$ profile, the surface temperature in this case is 328 K. Conversely, for the 100 bar case, we take the lower mass limit of 7.28 M$_{\oplus}$, obtaining a maximum ocean depth of 370 km. This is consistent with previous ocean depth estimates for the corresponding surface gravity\citep{Nixon2021}. In this case the surface lies at 398 K. In both cases, the total H$_2$O mass fraction in the interior is adjusted to minimise/maximise the ocean depth, with an upper limit of 90$\%$ \citep{Madhusudhan2021}. The H$_2$O adiabats, showing the temperature structures in the H$_2$O mantle, for both cases are shown in Figure \ref{fig:pt_profiles}. Therefore, based on our above estimates we nominally consider the ocean depths of our canonical Hycean world to range between 100-400 km, which we use in section~\ref{sec:nutrients}. 

\subsection{Atmospheric Composition}
Hycean worlds are characterised by their bulk atmospheric composition being dominated by H$_2$. However, the remaining molecular inventory of a Hycean atmosphere would depend on a number of factors, including the temperature structure, metallicity (or more generally, the bulk elemental abundances), surface pressure, the incident flux from the host star, potential atmospheric escape, and potential chemical fluxes to and from the ocean surface beneath the atmosphere. Many of these factors are unknown for Hycean worlds, particularly the initial conditions set by planet formation and the boundary conditions set by the ocean surface, which makes an ab initio modeling of their atmospheric chemistry challenging. Some recent studies have investigated the atmospheric chemistry of temperate sub-Neptunes with thin H$_2$-rich atmospheres for different assumptions about the surface and atmospheric parameters, which are also relevant for Hycean worlds \citep{Yu2021, Hu2021, Tsai2021}. Here we focus on our canonical Hycean atmosphere with temperature structures as discussed above and explore the possible atmospheric compositions for different assumptions about the initial and boundary conditions. 

As discussed in section~\ref{sec:methods_photochem}, we pursue a two-step approach. We first explore models assuming a chemically inactive (zero-flux) lower boundary to assess the sensitivity of the chemical processes to the assumed atmospheric properties, with no influence from the ocean, as pursued in recent work\citep{Yu2021,Tsai2021}. We then consider cases with fixed mixing ratio and deposition velocity boundary conditions for the ocean-atmosphere interface and assess their effect on the chemistry. We consider different types of boundary conditions based on expectations for a primordial atmosphere in thermochemical equilibrium or with significant CO$_2$ contribution from the ocean, as discussed in section~\ref{sec:methods_photochem}. We note that in this study we do not consider chemical fluxes resulting from potential life in the ocean which may significantly impact the atmospheric composition; we discuss this in section~\ref{sec:summary}.  

\begin{figure*}
    \centering
    \includegraphics[width=0.48\textwidth]{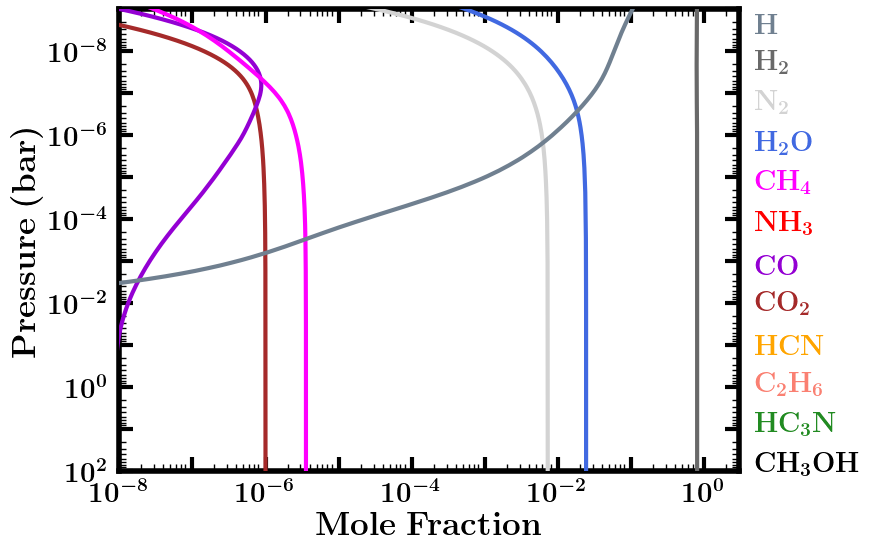}
    \includegraphics[width=0.48\textwidth]{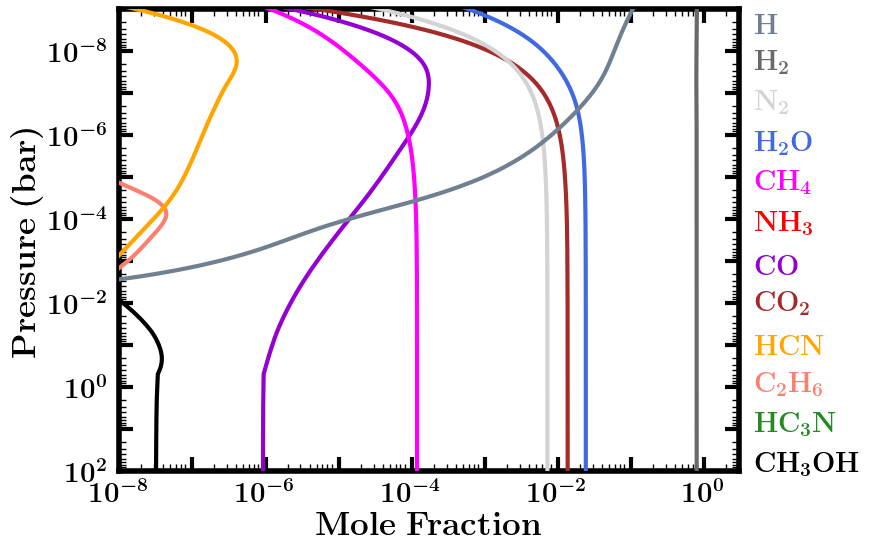}
    \caption{Steady-state mixing-ratio profiles considering an atmosphere with lower boundary conditions of an active ocean surface with  prescribed N$_2$, H$_2$O, and CO$_2$ mixing ratios and assuming the PT1 temperature profile in Figure~\ref{fig:pt_profiles}. Left: Abundances assuming a low CO$_2$ mixing ratio of 10$^{-6}$ at the atmosphere-ocean boundary, corresponding to Case 10 in Table~\ref{tab:chem_models}. Right: Abundances assuming a high CO$_2$ mixing ratio of $\sim$1.4$\times$10$^{-2}$ at the atmosphere-ocean boundary, corresponding to Case 9 in Table~\ref{tab:chem_models}.}
    \label{fig:pt1_3_sidebyside_co2}
\end{figure*}

Our results for the photospheric (1 mbar) abundances of prominent molecules for the different model scenarios discussed in section~\ref{sec:methods_photochem} are shown in Table~\ref{tab:chem_models}. The vertical abundance profiles for some representative cases are shown in Fig~\ref{fig:pt1_3_sidebyside}--\ref{fig:chemmodels}.
Fig~\ref{fig:pt1_3_sidebyside} shows the abundance profiles for our canonical model with zero-flux boundary conditions corresponding to Cases 1, 3, 4 and 6 in Table~\ref{tab:chem_models}. The abundances of the prominent molecules at 10 mbar and of other less abundant molecules are shown in the Supplementary Information. In what follows, we present our findings on the prominent C, N, O molecules in the model atmospheres.

\subsubsection{H$_2$ and H$_2$O} The most abundant molecules expected in a Hycean atmosphere are H$_2$, by definition, and H$_2$O, considering the ocean below. Even though H$_2$ itself does not have a strong spectral signature the 
H$_2$-rich nature of the atmosphere can be inferred via its low mean molecular weight, which in turn enables detections of other atmospheric constituents such as H$_2$O (\citep[e.g.][]{Benneke2019,Madhusudhan2020,Evans2023}). The H$_2$O abundance is governed by the temperature profile and atmospheric metallicity. For example, an atmospheric metallicity of 1$\times$ and 100$\times$ solar would imply an H$_2$O volume mixing ratio of $\sim$0.1\% and $\sim$10\%, respectively. If the abundance is above the local saturation vapour pressure, the excess H$_2$O can condense and precipitate out. Across our models with profiles PT1, PT2 and PT3, H$_2$O is abundant and is governed by saturation at the lower boundary. It is, therefore, generally expected that H$_2$O will be detectable in most Hycean atmospheres which are expected to be even warmer than K2-18~b \citep{Madhusudhan2021}. 

However, for the coolest Hycean worlds it is possible that the temperature at some point in the atmosphere is cold enough to lie within the ice sublimation regime causing H$_2$O to freeze out and further restricting the water vapor abundance at higher altitudes. This can be seen for our Case 3 and Case 6, using profiles PT3 and PT4, in Table~\ref{tab:chem_models} and in Fig.~\ref{fig:pt1_3_sidebyside}. This ``cold trap'' effect is seen both in the Earth's dry stratosphere and in the solar system giant planets where H$_2$O is negligible at the 0.1 bar level. In such cases it is possible that the H$_2$O may not be detectable in transmission spectra, which typically probe pressures below $\sim$0.1 bar. Thus, whether H$_2$O is observable or not depends on the temperature structure of the atmosphere, which in turn depends on the various factors discussed above. 

This potential cold-trapping effect is relevant to the ongoing debate regarding the observed transmission spectrum of the candidate Hycean planet K2-18 b, which has an insolation similar to that received by the Earth \citep{Benneke2019,Hardegree2020}. The transmission spectrum of the planet observed with HST has been explained by a degenerate set of solutions including H$_2$O and/or CH$_4$ \citep{Benneke2019,Tsiaras2019,Madhusudhan2020,Blain2021}. A non-detection or underabundance of H$_2$O in such an atmosphere could imply a cool and dry photosphere irrespective of the presence of an ocean underneath. This, in turn, could also imply the potential presence of H$_2$O clouds below the observable atmosphere. On the other hand, for hotter Hycean candidates an unambiguous detection of H$_2$O would be expected, as reported with the HST transmission spectrum of the planet TOI-270~d \citep{Evans2023}. Thus, the detection or non-detection of H$_2$O in a Hycean atmosphere can place important constraints on the Bond albedo of the planet, which strongly dictates the temperature structure in the observable upper atmosphere. It is useful to note that besides the temperature profile, H$_2$O is relatively unaffected by other factors in a Hycean atmosphere, e.g. photochemistry does not significantly alter the H$_2$O abundance.

\subsubsection{CH$_4$ and NH$_3$} 
In thick H$_2$-rich atmospheres, methane (CH$_4$) and ammonia (NH$_3$) are normally expected to be the dominant C and N bearing species under temperate conditions\citep{Lodders2002,Madhusudhan2012,Moses2013}. These molecules are ubiquitous in solar system giant planets. Both molecules are susceptible to photodissociation and other loss due to chemical kinetics in the observable atmosphere, but on giant planets are readily replenished due to their high abundances in the lower atmosphere, where higher temperatures maintain thermochemical equilibrium. However, the presence of a shallow surface boundary (whether ocean or solid) can significantly deplete the NH$_3$ and CH$_4$ abundances by curtailing their replenishment from the lower atmosphere \citep{Yu2021,Hu2021,Tsai2021}. In particular, we find that NH$_3$ is substantially depleted for surface pressures as high as 100 bar, even if no interaction with the surface occurs, as shown in Fig.~\ref{fig:pt1_3_sidebyside} and Table~\ref{tab:chem_models}. For comparison, for 100$\times$solar metallicity, the equilibrium abundance of NH$_3$ expected in our models is $1.5\times10^{-2}$. The lower the pressure at the surface, the greater the amount of NH$_3$ that is lost, as shown in Table~\ref{tab:chem_models} for Cases 4-6. These findings are consistent with previous studies \citep{Yu2021,Tsai2021}.

NH$_3$ is also highly soluble in liquid water, thereby increasing its depletion potential due to dissolution in the Hycean ocean, which acts as a sink.  This loss of NH$_3$ to the ocean is considered in our more realistic boundary condition models (\citep[see also][]{Hu2021}), where we find that the steady-state NH$_3$ mixing ratio becomes negligible; cases 9-11 in Table~\ref{tab:chem_models}. Therefore, a substantial depletion or a lack of NH$_3$ may be considered as a characteristic signature of a Hycean atmosphere. We note, however, the possibility that the presence of life in a Hycean ocean may itself be a source of NH$_3$, as has been suggested previously for habitable rocky exoplanets with H$_2$-rich atmospheres\citep{Seager2013,Ranjan2022}. 

\begin{figure*}
    \centering
    \includegraphics[width=0.48\textwidth]{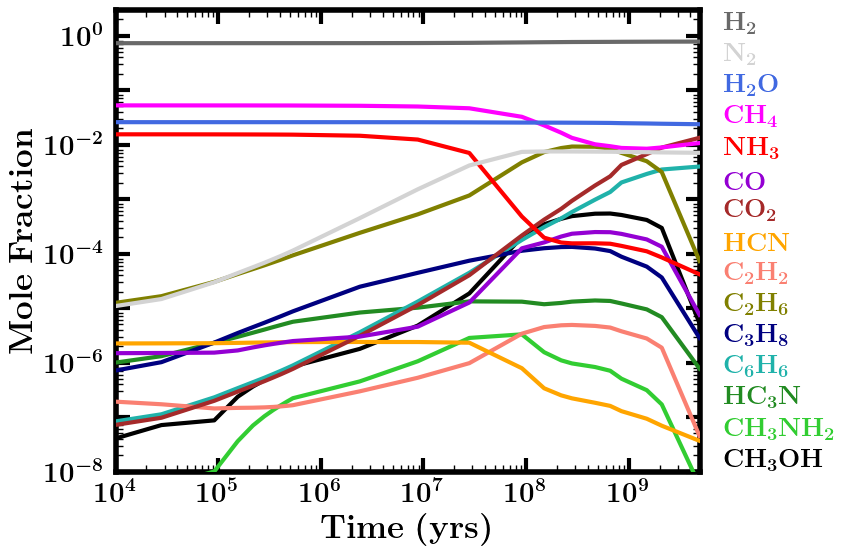}
    \includegraphics[width=0.5\textwidth]{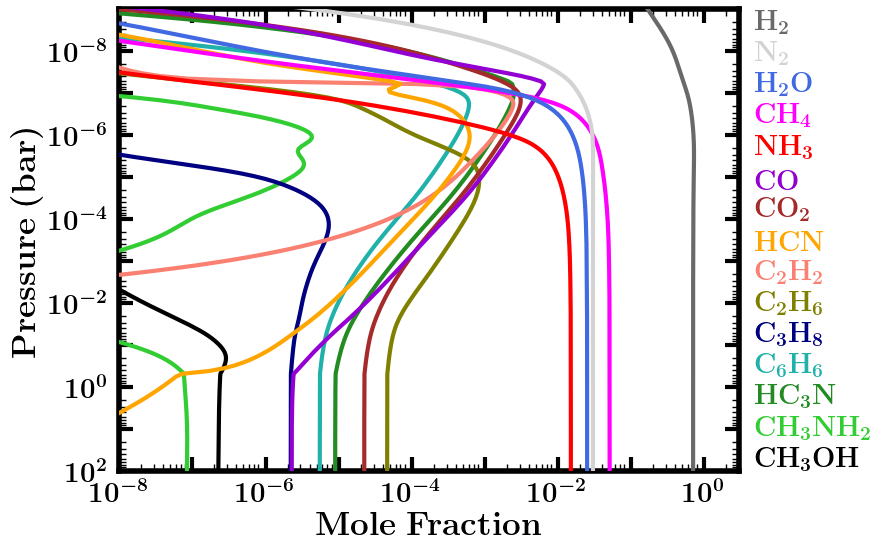}
    \caption{Left: Evolution of atmospheric molecular composition at 0.01 bar for Case 1 in Table~\ref{tab:chem_models}. Right: Abundance profiles corresponding to Case 7 in Table~\ref{tab:chem_models}, based on initial conditions of a reduced atmosphere in thermochemical equilibrium (similar to Case 1) and fixed boundary conditions for H$_2$O, CH$_4$ and NH$_3$ (see section~\ref{sec:methods_photochem}). The abundances reflect the composition at 28 Myr, showing the availability of organic species early in the evolution.}
    \label{fig:chemmodels}
\end{figure*}

The abundance of CH$_4$ is less discriminating compared to NH$_3$. Even though CH$_4$ is also susceptible to photodissociation and other chemical loss, it is expected to be more abundant than NH$_3$ due to multiple factors. Firstly, considering cosmic abundances\citep{Asplund2009}, C is more than twice as abundant as N, and CH$_4$ is more thermochemically stable than NH$_3$ at the warmer temperatures near the ocean surfaces of Hycean planets.
Secondly, in a H$_2$-rich atmosphere CH$_4$ is replenished by kinetic processes; that is, by photodissociation of other C bearing species such as CO and CO$_2$ in the upper atmosphere, and --- more significantly --- by recycling reactions deeper in the atmosphere, where higher pressures and temperatures allow photochemical products such as CO to be at least partially converted back to CH$_4$. We find significant CH$_4$ abundances for models evolved up to 5 Gyr for different surface pressures and metallicities, as shown in Table~\ref{tab:chem_models} and Fig.~\ref{fig:pt1_3_sidebyside}, consistent with previous studies\citep{Yu2021,Hu2021,Tsai2021}. We note that CH$_4$ is comparatively less abundant for lower metallicity and lower surface pressure, as also seen in previous work with zero-flux boundary conditions\citep{Yu2021}. 

Paradoxically, we find that warmer atmospheres in our studied range for 100-bar surface pressures retain the most CH$_4$, due to these aforementioned thermochemical recycling reactions being more efficient at higher temperatures. Atmospheres with ocean surfaces at lower pressures, on the other hand, lose more CH$_4$ as a result of missing out on these thermochemical reactions at greater depths, and they lose the methane on faster time scales. Depending on atmospheric temperatures, surface pressure, and planetary age, it is possible that CH$_4$ could be severely depleted and the C occupied primarily within CO$_2$ for atmospheres with high initial metallicities, shallow surfaces, older evolved compositions, or low carbon inventories (including low CO$_2$ dissolution scenarios, where CO$_2$ from the ocean is the only source of carbon). This latter scenario can be seen from our Cases 10 and 11 in Table~\ref{tab:chem_models} and Fig.~\ref{fig:pt1_3_sidebyside_co2}. Thus, a Hycean atmosphere might be expected to have observable amounts of CH$_4$ ($\gtrsim$10$^{-4}$) under most conditions, but several factors can affect those expectations. 

An important metric could be the relative abundance of CH$_4$ vs NH$_3$. If both molecules are present in high abundances consistent with expectations from thermochemical equilibrium, a Hycean ocean may be confidently ruled out. However, if NH$_3$ is absent or substantially depleted relative to CH$_4$ the presence of an ocean would be more likely. 

\subsubsection{Carbon Oxides: CO and CO$_2$}

In temperate and thick H$_2$-rich atmospheres, CO and CO$_2$ are generally expected through chemical disequilibrium processes and/or high metallicity. CO is expected through vertical mixing from a thick, deep atmosphere where high temperatures can thermochemically favor CO as the dominant C bearing species\citep{moses11}. On the other hand, CO$_2$ is expected in the presence of high-metallicity. Both CO and CO$_2$ can also result from C based photochemistry in the presence of a high H$_2$O abundance. However, the presence of an ocean underneath a thin H$_2$-rich atmosphere, as in a Hycean planet, can significantly enhance the CO$_2$ abundance further due to release of dissolved CO$_2$ from the ocean \citep{Hu2021}. Furthermore, over long timescales, as CH$_4$ is progressively photodissociated or otherwise chemically destroyed without replenishment from the lower atmosphere, CO$_2$ can become the dominant C bearing species in the atmosphere. We generally find a high CO$_2$ abundance across most of our cases, expecially those with the warmer $P-T$ profile (PT1) and 100$\times$solar metallicity, as shown in Table~\ref{tab:chem_models} and Fig.~\ref{fig:chemmodels}. CO is generally less abundant than CO$_2$ with the exception of cases where the overall oxygen abundance is low, due to freeze out of H$_2$O in the atmosphere, as in Case 6 for the coolest temperature profile we consider.

We note the important difference that the CO$_2$ abundance in our zero-flux boundary condition models is primarily due to chemical-kinetic processes in combination with initial metallicity assumptions, whereas that in Cases 9, 10, \& 11 and in the models of \citet{Hu2021} is an input to the model based on the assumption for how much CO$_2$ is released from the ocean. In the \citet{Hu2021} models, the abundance of atmospheric CO$_2$ is controlled by whatever fixed mixing ratio they adopt for CO$_2$ at the lower boundary, which they nominally assume to be between 4$\times$10$^{-4}$ and 0.1; their estimate for the lower-bound ranges between 5$\times$10$^{-5}$ -- 7$\times$10$^{-4}$ bar partial pressure of CO$_2$. We also examine cases where the only carbon present in the atmosphere is that released through equilibrium with CO$_2$ dissolved in the ocean, with an assumed CO$_2$ mixing ratio at the ocean surface of 10$^{-6}$ (i.e., CO$_2$ partial pressure of 10$^{-4}$ bar for our 100-bar atmosphere case), $\sim$1.4$\%$), or 10$\%$. These correspond to cases 9-11 in Table~\ref{tab:chem_models} and Fig.~\ref{fig:pt1_3_sidebyside_co2}. 

As with the \citet{Hu2021} study, we find that adopting fixed mixing-ratio lower-boundary conditions for CO$_2$, N$_2$, and H$_2$O in a canonical ocean model results in an atmosphere with much larger CO$_2$/CO ratios than for otherwise similar planets without ocean interaction.  We agree that the CO$_2$/CO ratio could be a useful indicator to distinguish between high-metallicity deep atmospheres and planets with shallower atmospheres and liquid-water oceans. However, our high CO$_2$ ocean-release cases produce much less CH$_4$ than is derived for the \citet{Hu2021} models, which end up with more CH$_4$ than CO$_2$, despite CO$_2$ being the only incoming source of carbon in the model.

While the presence of substantial atmospheric CO$_2$ and high CO$_2$/CO ratios may be  strong indicators of a Hycean ocean, the  absence of CO$_2$ --- while less likely on a warm Hycean planet --- might also be less informative. For example, following the approach of \citet{Hu2021} a low CO$_2$ abundance could indicate (a) a high efficiency of dissolution of CO$_2$ in the ocean and less availability in the atmosphere, or (b) more C sequestered in the core, making it less available in the ocean, or a generally low C content in the planetary interior as a whole. Secondly, the abundance of CO$_2$ in the atmosphere can also depend on the initial availability of CH$_4$ and H$_2$O in the atmosphere through the CH$_4$ + H$_2$O $\rightarrow$ CO$_2$ conversion process (i.e., assuming that the primordial atmosphere contained CH$_4$ not in immediate chemical equilibrium with the liquid ocean, such that the oceanic CO$_2$ does not control the entire atmospheric carbon budget). For cooler Hycean planets where H$_2$O can be frozen out in the observable atmosphere, CO$_2$ could also be underabundant from this process and hence less observable. Finally, it is also possible that microorganisms in a Hycean ocean could efficiently use dissolved CO$_2$ for biological processes and subsequently release other gases such as CH$_4$ or O$_2$. Microbial methanogenesis is known to be a significant source of CO$_2$ consumption in the Earth's oceans \citep{Hoehler2001,Bains2014,Berghuis2019} and it is not inconceivable that life in a Hycean ocean may find efficient ways to do the same. 

\subsubsection{Other Molecules} Most of our models contain moderate (typically ppb to ppm) levels of HCN, but because HCN is produced from photochemistry of both N$_2$ and NH$_3$ in the presence of other carbon-bearing molecules, it does not seem to be a good indicator of the presence/absence of an ocean.  \citet{Tsai2021} also emphasize that HCN is very sensitive to stellar properties and the strength of atmospheric mixing, so HCN becomes less reliable than NH$_3$ as an indicator of the presence of a surface.  We do find that HCN is more depleted when a sink at the oceanic lower boundary is included, as in Cases 9-11.

Similarly, CH$_3$OH becomes very abundant in some of our zero-flux boundary condition cases (e.g. Cases 2 \& 3 in Table~\ref{tab:chem_models} and Fig.~\ref{fig:pt1_3_sidebyside}) but is more depleted in our models with ocean interaction, due to dissolution in the ocean.  In that way, CH$_3$OH can become a good discriminator of ocean versus solid surfaces, as suggested originally by \citet{Tsai2021}.  Note also that the CH$_3$OH abundance in our zero-flux models has a high sensitivity to atmospheric temperatures, being less abundant in some atmospheres because it becomes a casualty of the more efficient conversion of CO into CH$_4$ at depth in warmer atmospheres (as in Cases 1 and 4, see e.g. Fig.~\ref{fig:pt1_3_sidebyside}), or becomes a casualty of the low atmospheric oxygen abundance that results from H$_2$O freeze out in colder atmospheres (as in Case 6, e.g., Fig.~\ref{fig:pt1_3_sidebyside}).

With our cooler zero-flux models with a 100-bar surface, we find that C$_2$H$_6$ could also be an excellent indicator of a dry, unreactive planetary surface on a sub-Neptune planet. This can be seen in Fig.~\ref{fig:pt1_3_sidebyside}. If H$_2$O is moderately depleted through condensation/sublimation near a tropopause cold trap (e.g., with our PT2 \& PT3 profiles, Cases 2-3), such that H$_2$O has a reduced abundance in the middle and upper atmosphere, conversion of CH$_4$ into CO and CO$_2$ becomes less effective than conversion of CH$_4$ into heavier hydrocarbons. In that situation, the overall photochemistry becomes more like that of Jupiter, except the hydrocarbons are not recycled efficiently back to CH$_4$ at depth.  The eventual steady-state abundance of photochemically stable hydrocarbons such as C$_2$H$_6$ and other alkanes (e.g., C$_3$H$_8$ and C$_4$H$_{10}$) becomes very large in cases with no surface loss, such that C$_2$H$_6$ actually contains the bulk of the carbon in the atmosphere.  We are also finding a significant C$_6$H$_6$ abundance in our zero-flux models, but that result seems less robust, as we end the carbon chemistry as C$_6$H$_6$ and do not consider further loss of benzene to PAHs and other heavy organics.  In our ocean models with fixed CO$_2$, N$_2$, and H$_2$O boundary conditions, we assume a moderate deposition velocity loss of C$_2$H$_6$ through the ocean boundary, and the steady-state C$_2$H$_6$ mixing ratio is less dramatic.  Therefore, C$_2$H$_6$ should provide a better observational marker than CH$_3$OH to distinguish ocean planets from solid-surface planets on sub-Neptunes that are cold enough for significant atmospheric water depletion. Another species of potential note is HC$_3$N, which becomes relatively abundant for some of our models.

\begin{figure*}
    \centering
    \includegraphics[width=\textwidth]{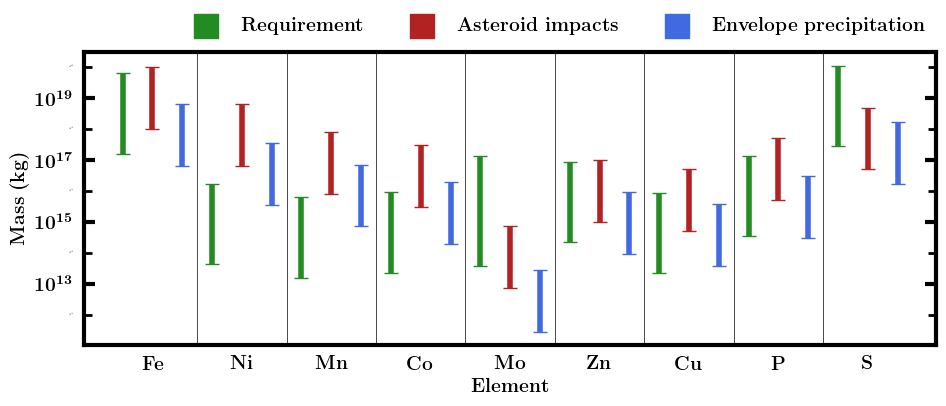}
    \caption{Inventory of prominent bioessential elements. The estimated masses of the elements required for our canonical ocean are shown in green, along with the masses that can be available from impacts (brown) and from atmospheric settling (blue). See section~\ref{sec:nutrients}.}
    \label{fig:nutrient_mass_overview}
\end{figure*}

\subsection{Prebiotic Molecular Inventory}
The molecular inventory of a habitable planet in the early stages of its evolution is expected to play an important role in its propensity for seeding life. On the Earth, key pre-biotic molecules such as HCN are thought to have played a pivotal role in abiogenesis \citep{Rimmer2018}. Here, we investigate the molecular abundances over the lifetime of our canonical Hycean planet. At the outset, we find that most of the models considered undergo a phase that is rich in organic molecules early in the evolution, typically within the first $\sim$1-300 Myr. The relative abundances of the different molecules depend on the boundary conditions and other factors discussed above. However, a common outcome in our zero-flux models is the prevalence of important organics such as CH$_4$, C$_2$H$_6$, C$_6$H$_6$, and nitriles over the early evolutionary history, as shown in Fig~\ref{fig:chemmodels}, left panel.  

We find that our models with fixed boundary conditions corresponding to an initially reduced atmosphere result in abundant organics starting very early in the evolution. We explored this in Cases 7 and 8 in Table~\ref{tab:chem_models}; the abundance profiles for Case 7 are shown in Fig~\ref{fig:chemmodels}, right panel. We find that hydrocarbons, nitriles, alcohols, and other organics can become extremely abundant in the first few million years of atmospheric evolution, if we assume the original starting composition were reduced (e.g., CH$_4$, NH$_3$, and H$_2$O being the main carriers of C, N, and O, as would be expected for a cool H$_2$-rich accretion atmosphere in thermochemical equilibrium in the gaseous envelope, but not the ocean); see also \cite{Yu2021,Tsai2021}.  High abundances of C$_2$H$_6$, C$_3$H$_8$, C$_4$H$_{10}$, C$_6$H$_6$, C$_2$H$_2$, HCN, H$_3$CN, CH$_3$NH$_2$, CH$_3$CN, and CH$_3$OH are particularly worth mentioning from our model results.  While it may be unlikely that we manage to catch this early evolutionary stage in observations of Hycean planets, these photochemically produced molecules could provide pre-biotic molecules to the ocean. The abundance of HCN, an important prebiotic molecule on Earth\citep{Rimmer2018}, is strongly correlated with the presence of NH$_3$ in the atmosphere. As NH$_3$ is photochemically destroyed over time, and is readily soluble in the ocean, most of the N is locked in N$_2$ thereby reducing the abundance of HCN at later times. However, other hydrocarbons with lesser or no dependence on N continue to be abundant. Heavier organic species are present but not particularly abundant in our final steady-state atmospheres with ocean interaction included, due to relatively low abundances of the primary parent molecule CH$_4$. However, if CH$_4$ continues to be abundant in the present-day atmosphere due to the reasons described above, a rich organic molecular inventory may continue as well.

\subsection{Availability of Bioessential Elements}
\label{sec:nutrients}
 A central challenge to the habitability of Hycean worlds, as for all ocean worlds, is the availability of bioessential elements. Bioessential elements include the common volatile elements (C H N O P S) as well as essential metals such as Fe, Na, K, Mg, Mn, Ni, Ca, Cu, Zn, Co, and Mo \citep{Cockell2016,Hendrix2019, Robbins2016, Seager2021}, which perform key functions in biochemical processes in terrestrial life. The difficulty lies in the fact that ocean worlds would not have ready access to these  elements, which are typically acquired through interaction of the hydrological cycle and the silicate portion of the planet. This problem is particularly acute for planets with deep oceans where high-pressure ice at the bottom of the liquid water column can preclude access to elements from the rocky core. This challenge is well recognised both for volatile-rich sub-Neptunes \citep{Seager2021,Cockell2016} as well as for icy moons in the solar system\citep{Hendrix2019}. For planets with H$_2$O-rich interiors and H$_2$-rich atmospheres, including Hycean worlds, H and O are naturally present in abundance while C and N are also found in large quantities in volatile ices (e.g. \citet{Rubin2019}) and so could be present in the required abundances. However, this is not necessarily the case for P, S, and refractory metals. Here we assess possible sources of these elements for our canonical Hycean world. 

\subsubsection{Mass Requirements}
We first estimate the elemental inventory required for life in a canonical Hycean ocean assuming similar concentrations as those found in the Early Earth \citep{Robbins2016, Anbar2008, Crowe2014, Jones2015}. We consider ocean depths of 100-400 km for a canonical Hycean world similar to K2-18~b, with $R_p =$ 2.61 R$_\oplus$. We note that here we are primarily concerned with the amount of nutrients required to seed life on the canonical Hycean as in the oceans of the Early Earth. As we know Early Earth ocean elemental concentrations and the volume of our Hycean ocean,  we calculate the total Hycean mass requirements such that elemental concentrations (in Mol / L) in these two oceans match. 

We use \citet{Robbins2016} estimates of Archean ocean metal concentrations. These originate from geochemical modelling by \citet{Saito2003}, except for the Mo values which are taken from \citet{Anbar2002}. We take \citet{Jones2015} and \citet{Crowe2014} values for the P and S concentrations respectively. These estimates are obtained through a variety of methods. The \citet{Saito2003} results come from geochemical modelling of the concentrations of various metal species under anoxic and iron-rich conditions representative of Archean oceans\citep{Robbins2016, Anbar2008, Swanner2020, Saito2003, Tosca2016}.The modelling finds that the differential availability of various metals follows Fe$\, >\, $Ni, Mn, Co$\, \gg \, $Zn, Cu. It is assumed that there is sufficient nutrient influx (e.g. from weathering) for the actual earth ocean concentrations to match these modelling estimates. The P, S, and Mo values, in contrast, are taken from proxy sedimentary records. This can offer a more nuanced view of changes over time but can be limited by imperfect sorption models and post-deposition alteration of the sediments \citep{Robbins2013, Robbins2016, Mukherjee2020}.

There is one more step to be made before arriving at the estimated mass requirements. A proportion of the elemental mass budget will be present in a precipitate instead of a dissolved form. Since the former will not be bioavailable, we adopt a dissolved fraction\citep{Saito2003, Anbar2002, Crowe2014, Jones2015} to scale our Hycean mass budget for each element. This increases the required mass budget for each element. We note that we have not modelled self-consistent Hycean ocean chemistry or any kind of (geo)chemical cycle and so there is still significant uncertainty in this dissolved fraction. The Hycean ocean considered here would be warmer than the earth ocean, meaning that the solubilities of various species would be different than on Earth.  To reflect these unknowns, we introduce uncertainties of 1 dex in the primitive ocean concentrations, except for Mo for which we consider an increased uncertainty of 1.5 dex as its Archean abundance is relatively less accessible. The mass requirements are shown in Table \ref{tbl:nutrient_req_asteroids} and Figure \ref{fig:nutrient_mass_overview}. 

 \subsubsection{External Delivery}
 One of the most common sources of elemental delivery in planetary environments is through external impacts by asteroids over the early history of the planet. Considering representative impactors on the Hadean Earth, we estimate the concentrations of nutrients that can be delivered to the Hycean ocean. We assume a composition of carbonaceous chondrites \citep{Lodders2020}, which are somewhat lower in metal content relative to other asteroids. For the impacting masses, we take the minimum estimate of post Moon-formation impacts \citep{Zahnle2020}, itself derived from the crater size distribution on the Moon. This gives a total asteroid mass of $2\times10^{21}$kg which impacted the Earth, which we set as the total asteroid mass impacting the Hycean world. The mass distribution is weighted towards larger objects, $M(<m)\propto m^{0.5}$\citep{Zahnle2020}. While the estimates for the total mass delivered to Earth in this Late Heavy Bombardment / Late Veneer scenario vary (as does the precise timing of these events, although this is not relevant to us), our value is on the lower end of usual estimates \citep{Parkos2018, Nesvorny2023}.

Only a small fraction of the $2\times10^{21}$kg will end up dissolved in the Hycean oceans. Large impacts can travel through the ocean and deliver mass to the ice layers or even the core. Hydrodynamic modelling of asteroid impacts into Earth's oceans \citep{Gisler2011, Davison2007, Nishizawa2020} suggests impactors will form impact craters if the ocean depth is less than 5-7 times the impactor diameter. If this factor applies to our deep hycean oceans, only objects with a diameter < $\sim28$ km would break apart completely. Taking a chondritic asteroid density of $1600$ kg $m^{-3}$ \citep{Carry2012} this would mean only objects lighter than $\sim2\times10^{16}$ kg, and so only $\sim0.5\%$ of the total mass would be in impactors small enough to completely break apart in the ocean.
Larger collisions that do reach a solid surface will throw up material back into the atmosphere, which will rain back down as small particles \citep{Genda2017, Parkos2018}. On the Earth, large collisions excavate material with a total mass nearly as large as the impacting mass \citep{Johnson2012}, significantly increasing the amount of mass deposited on the planet surface. 

However, as little work on large impacts in deep oceans has been done, we are cautious about this additional amount of mass delivered to the oceans. We assume that between $0.25\%$ and $25\%$ of the total impact mass stays in the oceans, with a median of $1\%$ or $2\times10^{19}$kg, and provides the required source of bioessential elements for seeding life. This fraction will change depending both on ocean depth and the precise physics of impact. The mass requirements and asteroid nutrient delivery rates are shown in Table \ref{tbl:nutrient_req_asteroids} as well as Figure \ref{fig:nutrient_mass_overview}. We find that in general asteroid impacts can plausibly deliver enough mass for the bioessential elements with the possible exception of Molybdenum and Sulphur.
\begin{table}[h]
\centering
\begin{tabular}{cccccc}
\hline \hline
Element & $C_{\rm PO}$ & $M_{\rm req}$ & $f_{\rm CC}$ & $F_{\rm diss} $ & $M_{\rm imp}$ \\
\hline
Fe & -4.6 & 3.2E18 & 2.0E-01 & 3.0E-01 & 4.0E17\\
Ni & -8.2 & 8.5E14 & 1.3E-02 & 3.0E-01 & 2.6E17\\
Mn & -8.3 & 3.2E14 & 1.6E-03 & 6.0E-01 & 3.2E16\\
Co & -8.1 & 4.6E14 & 6.0E-04 & 7.0E-01 & 1.2E16\\
Mo & -9.0 & 2.1E15 &   1.5E-06 & 3.0E-02 & 3.0E13\\
Zn & -14.0 & 4.5E15 & 2.0E-04 & 1.0E-07 & 4.0E15\\
Cu & -22.0 & 4.4E14 & 1.0E-04 & 1.0E-14 & 2.0E15\\
P & -6.5 & 6.9E15 & 1.0E-03 & 9.8E-01 & 2.0E16\\
S & -5.0 & 5.7E18 & 1.0E-02 & 4.0E-02 & 2.0E17\\
\hline
\end{tabular}
\caption{Estimates of element masses required to provide the necessary concentrations in our canonical Hycean ocean. $C_{PO}$  is the estimated  concentration of a given element in earth's primeval ocean\citep{Robbins2016}, shown in $\log({\rm moles/litre})$, and $M_{\rm req}$ is the median estimate of the element mass (in kg) required, assuming a median ocean depth of 200 km. The estimated range is shown in Fig.~\ref{fig:nutrient_mass_overview}. $f_{CC}$ refers to our adopted mass fraction of the element in carbonaceous chondrites, and $F_{diss}$ is the solubility fraction of the element dissolved in Archean ocean conditions on Earth\citep{Saito2003,Anbar2002}. $M_{\rm imp}$ is the median estimate for the mass (in kg) of that element delivered from asteroid impacts.}\label{tbl:nutrient_req_asteroids}
\end{table}

Another avenue for nutrient delivery is through steady state accretion of extraterrestrial dust. The possibility of extraterrestrial dust being a sustained source of bioessential nutrients has been considered in the context of both the Earth \citep{Reiners2018} and sub-Neptunes \citep{Seager2021}. For example, in the context of the Earth, the amount of Fe obtained through this source is estimated to be $4\times10^6$ kg/yr \citep{Peucker2016}, compared to a global phytoplankton Fe assimilation of $7\times10^8$ kg/yr \citep{Fung2000}. So this flux would only be able to support a small biosphere by itself, and it would take of order $10^{12}$ yr for it to raise the overall ocean Fe budget to the required $\sim10^{18}$ kg. While it may help life continue once it has started, it may not be a substantial contributing factor to the start of life, unless the dust accretion rates are significantly higher at early times. On the other hand, depending on the system parameters the impactor and dust influx on Hycean worlds may be significantly higher than assumed here based on Earth values. Future work can investigate if the larger planet sizes of Hycean worlds and their closer proximity to their host stars, among other system properties, may enhance the mass influx rate. 

\subsubsection{Atmospheric Condensation}
 Another important source of nutrients in a Hycean world would be the initial atmospheric composition. Depending on the formation scenario, the primordial hydrogen-rich atmosphere of a Hycean planet can contain a significant abundance of metals reflecting their proportions in the H-rich nebula. For example, a solar composition gas contains $0.13\%$ by mass in Fe, $0.007\%$ in Ni, etc. \citep{Lodders2020}, as shown in Table~\ref{tbl:atm_mass}. However, this abundance is not a given considering that the planet could form outside the metal condensation fronts in the protoplanetary disk and accrete metal-poor gas. On the other hand, if the planet forms closer in or accretes significant solids along with the volatiles during formation it is conceivable that an equivalent amount of metals to a nebular metallicity gas, or even higher metallicity, may be accreted and accessible in the volatile layer. If initially present in the atmosphere, these metals would eventually condense out given the temperate atmosphere of a Hycean world. We estimate whether such a scenario could meet the bioessential metal requirement in a Hycean ocean. 

 We consider a nominal atmospheric mass of our canonical Hycean planet to range between 10$^{-6}$ - 10$^{-4}$ M$_p$, informed by our internal structure model and previous estimates for K2-18~b\citep{Madhusudhan2020}. Considering a solar composition for the atmosphere, we estimate the elemental budget of key metal species, as shown in Table \ref{tbl:atm_mass}. For P and S, which are not expected to condense out in the same fashion, we take a Neptune-like 100x solar metallicity atmosphere and assume that $1\%$ of the atmospheric species end up in the ocean, very likely an underestimate \citep{Hu2021}. We find that condensation in our primordial atmosphere provides comparable elemental inventory to the ocean requirements. Figure \ref{fig:nutrient_mass_overview} shows these estimates alongside the masses of required nutrients and possible asteroid impacts. 

\begin{table}[h]
\centering
\begin{tabular}{ccccc}
\hline \hline
Element & X/H & $f_{\rm atm}$ & $M_{\rm req}$ & $M_{\rm med}$ \\
\hline
Fe & 3.0E-05 & 1.3E-03 & 3.2E18 & 6.6E17 \\
Ni & 1.6E-06 & 7.2E-05 & 8.5E14 & 3.7E16 \\
Mn & 3.3E-07 & 1.1E-05 & 3.2E14 & 7.1E15 \\
Co & 8.5E-08 & 4.2E-06 & 4.6E14 & 2.0E15 \\
Mo & 7.6E-11 & 5.4E-09 & 2.1E15 & 2.9E12 \\
Zn & 3.6E-08 & 1.7E-06 & 4.5E15 & 9.3E14 \\
Cu & 1.5E-08 & 7.2E-07 & 4.4E15 & 3.8E14 \\
P & 2.6E-07 & 5.9E-06 & 6.9E15 & 3.1E15\\
S & 1.3E-05 & 3.2E-04 & 5.7E18 & 1.7E16\\
\hline
\end{tabular}
\caption{Estimation of mass budget of trace metals contained in the atmosphere. X/H is the number fraction of the element relative to H in solar abundance\citep{Lodders2020}. $f_{\rm atm}$ is the atmospheric mass fraction in a given element for a solar composition atmosphere. $M_{\rm med}$ is the median estimate mass (in kg) of the element delivered to the ocean; the estimated range is shown in Fig.~\ref{fig:nutrient_mass_overview}. }\label{tbl:atm_mass}
\end{table}

\subsubsection{Convective Transport from Core}
The lack of an ocean-rock boundary, preventing the dissolution of key nutrients and minerals, has been a longstanding objection to life both on ocean exoplanets and larger icy moons such as Ganymede \citep{Lingam2018,Journaux2020,Noack2016}. However, evidence is building that even on planets with thick high-pressure ice layers, there may still be transport within the ice layer through convection \citep{Choblet2017, KS2018, Kalousova2018, Lebec2023}. \citet{Hernandez2022} found that ice VII could transport up to $2.5\%$ NaCl in its crystalline structure, although it is not clear what the situation for other nutrients would be. A number of unknowns prevent us from estimating the nutrient mass budget in the ocean possible through this mechanism. In particular, we expect ice X near the Hycean core boundary, and its behaviour is even less understood than ice VII. However, some form of nutrient transfer from the core to the ocean may be plausible.

 \begin{figure*}[ht]
    \centering
    \includegraphics[width=\textwidth]{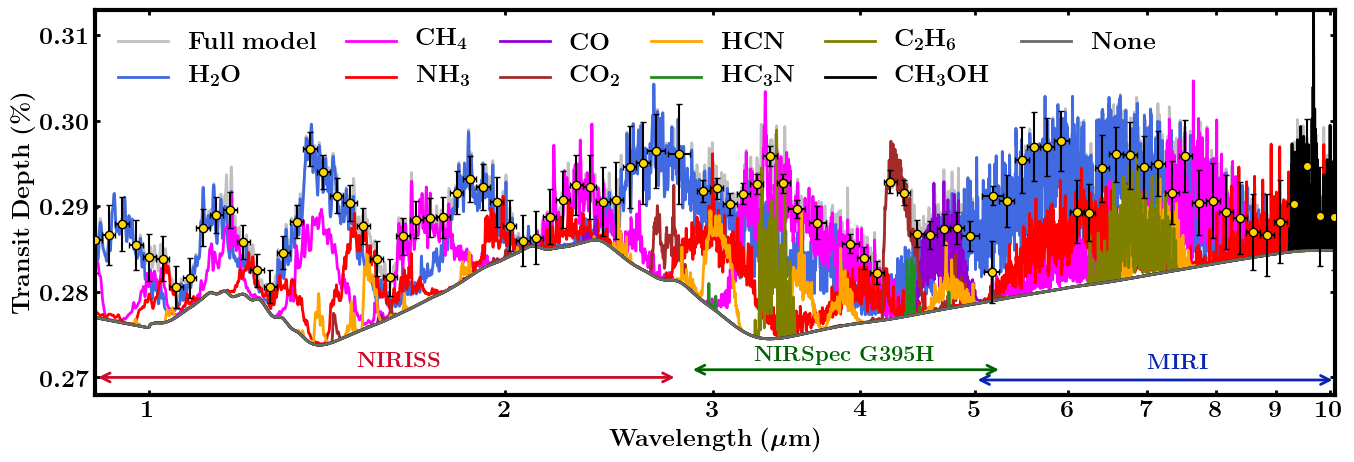}
    \caption{Model transmission spectrum of our canonical Hycean world K2-18~b. The model spectrum with all the molecules included is shown in grey. The spectra with the individual contributions of the different molecular species are shown as denoted in the legend. The synthetic JWST observations corresponding to the full model are shown as the gold-centered black points with error bars binned to a resolution of R$=40$. The observations include one transit with NIRISS SOSS, three transits with NIRSpec G395H and two transits with MIRI LRS, as discussed in section~\ref{sec:jwst}.}
    \label{fig:spectra}
\end{figure*}

 \begin{figure*}[ht]
    \centering
    \includegraphics[width=\textwidth]{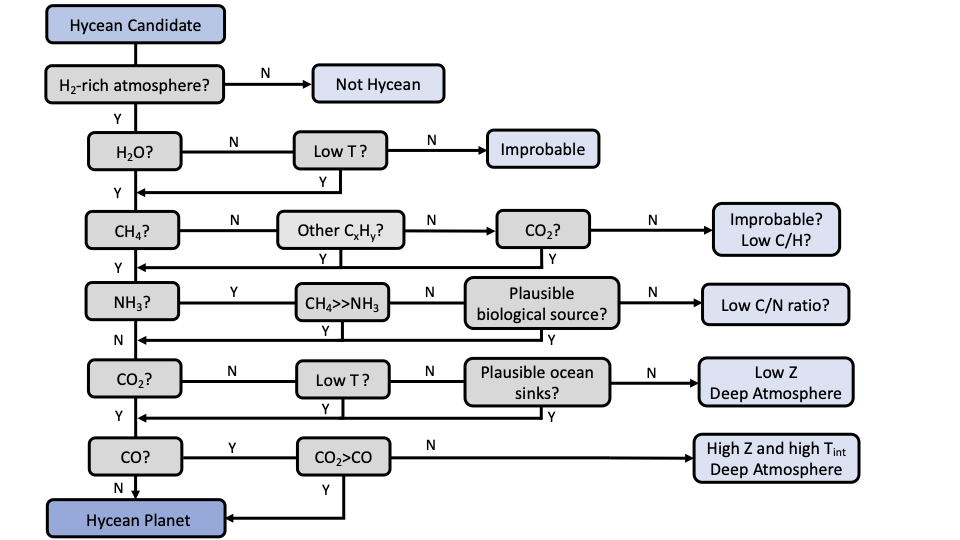}
    \caption{Chemical diagnostics for Hycean atmospheres based on our photochemical models and motivated by recent works\citep{Yu2021,Hu2021,Tsai2021}. The relevance of each prominent molecule for the identification of a Hycean atmosphere are shown. For each molelcule, the implications of its detection or non-detection are indicated, considering a minimal  detectability threshold of 1 ppm abundance. Low T refers to cases where the atmospheric temperature is expected to be cold enough for H$_2$O condensation. C$_{\rm x}$H$_{\rm y}$ refer to hydrocarbons. Z refers to atmospheric metallicity in a deep H$_2$-rich atmosphere, and T$_{\rm int}$ refers to the internal temperature. The pathways reflect the results of our photochemical calculations across the range of Hycean conditions considered and are generally consistent with previous works\citep{Yu2021, Hu2021, Tsai2021}. See section~\ref{sec:jwst}.}
    \label{fig:flowchart}
\end{figure*}

\section{Observational Prospects with JWST}
\label{sec:jwst}

A central promise of Hycean planets is their accessibility to atmospheric characterisation with current facilities, particularly the JWST \cite{Madhusudhan2021}. Their low gravities and low atmospheric mean molecular weight make Hycean worlds more conducive to atmospheric spectroscopy compared to rocky exoplanets of comparable mass and temperature. Several recent studies have demonstrated the feasibility of detecting  key molecular signatures, including biomarkers, in candidate Hycean worlds using transit spectroscopy with JWST\citep{Madhusudhan2021,Constantinou2022,Hu2021}. Our goal here is to assess (a) the spectral features of relevant chemical species observable with JWST transit spectroscopy, and (b) chemical diagnostics for identifying a Hycean planet. 

\subsection{Atmospheric Spectroscopy with JWST}
The high sensitivity and broad spectral range of JWST provide a promising avenue to detect key molecular species expected in Hycean atmospheres using transit spectroscopy. Here we assess the spectral features of the molecules predicted in our photochemical calculations, in section \ref{sec:results}, that may be observable with JWST. We model a transmission spectrum of K2-18~b using an adaptation of the AURA exoplanet atmospheric modeling and retrieval code\cite{Pinhas2018}. We refer the reader to previous works\cite{Madhusudhan2021,Constantinou2022} for details on the model setup using the AURA code and for the same planet bulk parameters as considered here. The model considers a plane parallel atmosphere at the day-night terminator of the planet in hydrostatic equilibrium and computes the transmission spectrum for a given temperature structure and composition. Besides the molecules considered in previous works using the AURA model \cite{Pinhas2018,Madhusudhan2021,Constantinou2022}, we consider three additional molecules C$_2$H$_6$\cite{HITRAN2020,C2H6-nu-12-1425,H-658}, HC$_3$N\cite{HITRAN2020,C2H6-nu-12-1425,HC3N-gamma_air-1-614}, and CH$_3$OH\cite{HITRAN2020,C2H6-nu-12-1425,CH3OH-gamma_air-1-568, CH3OH-S-2-569}. As seen in section~\ref{sec:results}, the atmospheric composition can vary widely depending on a number of factors, including the temperature structure, atmospheric metallicity, ocean surface pressure, and ocean-atmosphere interactions. We consider an isothermal temperature structure at 300 K and uniform volume mixing ratios of 10$^{-2}$ for H$_2$O, 10$^{-4}$ for CH$_4$, and 10$^{-5}$ for all other species. While these values lie within the wide range of abundances predicted by our chemical models, we note that some of the species may be significantly less or more abundant in a Hycean atmosphere than assumed here for illustration. 

The model transmission spectrum along with the relative contributions of the molecules and synthetic JWST observations are shown in Fig.~\ref{fig:spectra}. Following previous work\citep{Madhusudhan2021}, we assume 1 transit with JWST NIRISS\citep{Doyon2012} and 3 transits with NIRSpec G395H\citep{ferruit2012, Birkmann2014}, and an additional 2 transits with the MIRI\cite{Rieke_2015} spectrograph to obtain a broad 1-10 $\mu$m spectral coverage. The simulated spectra were generated using Pandexo\citep{Batalha2017}. The resultant five transits of K2-18~b can be observed in $\sim$50 hours with JWST. As can be seen in Fig.~\ref{fig:spectra}, and consistent with previous studies\citep{Madhusudhan2021,Constantinou2022,Hu2021,Tsai2021}, the prominent molecules H$_2$O, CH$_4$, NH$_3$, CO$_2$ and CO have strong  spectral features across the 1-10$\mu$m range and are readily discernible with the expected JWST data quality if present at the atmospheric abundances assumed in the model. Conversely, if any of them are underabundant then robust upper-limits can be placed on their abundances. For example, simulated retrieval studies\citep{Madhusudhan2021,Constantinou2022} have shown that the abundances of H$_2$O, CH$_4$, and NH$_3$ can be retrieved to precision better than 0.3 dex with JWST quality data. Among the minor species, HCN and C$_2$H$_6$ have the stronger features, whereas CH$_3$OH and HC$_3$N would be more challenging to detect, as previously reported for CH$_3$OH\cite{Tsai2021}. 

\subsection{Chemical Diagnostics of a Hycean World}

As discussed in section~\ref{sec:intro}, chemical characterisation of the atmosphere is essential to identify a Hycean world. The bulk properties of a candidate Hycean world, i.e. the mass, radius, and temperature, can generally be consistent with a degenerate set of internal structures\citep{Madhusudhan2020}, as shown in Fig.~\ref{fig:subneptunetypes}. Depending on the specific bulk properties, the possibilities could include (a) a Hycean world with a shallow H$_2$-rich atmosphere overlying a habitable ocean, (b) a sub-Neptune with a shallow H$_2$-rich atmosphere but with a solid surface underneath, or (c) a sub-Neptune with a deep H$_2$-rich atmosphere with temperatures and pressures at the bottom of the H$_2$ layer too high to be habitable. Several recent studies have investigated the possibility of atmospheric chemical signatures to identify the presence of solid or ocean surfaces in sub-Neptunes\citep{Yu2021,Hu2021,Tsai2021}, considering the possibility of a shallow H$_2$-rich atmosphere\citep{Madhusudhan2020} in K2-18~b as a canonical example. Here we revisit this avenue based on the chemical calculations in the present work and motivated by previous works\citep{Yu2021,Hu2021,Tsai2021}. A flowchart of the chemical diagnostics of a Hycean world is shown in Fig.~\ref{fig:flowchart}. 

Given a candidate Hycean world\citep{Madhusudhan2021}, the first requirement is to confirm the presence of an H$_2$-rich atmosphere using spectroscopic observations. An H$_2$-rich atmosphere is usually inferred indirectly through the detection of a spectral feature of one of the prominent molecules, e.g. H$_2$O, the spectral amplitude of which provides a constraint on the mean molecular mass of the atmosphere\citep{Benneke2019,Evans2023}. As discussed above, JWST will be able to robustly detect spectral features and constrain H$_2$-rich atmospheres for several nearby Hycean planets.

H$_2$O is naturally expected to be the dominant molecule observable in a Hycean atmosphere. However, if the temperature structure is too cool in the observable photosphere H$_2$O can be depleted due to condensation, as in the Earth's stratosphere. As discussed in section~\ref{sec:results}, for some of the $P$-$T$ profiles of K2-18~b considered, H$_2$O can indeed be underabundant. On the other hand, warmer Hycean candidates would be expected to have discernible H$_2$O features. 

The abundance of CH$_4$ in a Hycean atmosphere depends strongly on the boundary conditions. While CH$_4$ is susceptible to loss from photochemical processes, especially for low ocean surface pressures (e.g. 1 bar), it can still be present in significant abundances due to other production pathways. The underabundance of CH$_4$ would normally indicate the C present in other molecules such as CO$_2$ or, in the case of low-temperature Hyceans, higher-order hydrocarbons. The simultaneous depletion of all these molecules would be unlikely for a Hycean atmosphere, and may instead indicate a deep atmosphere with an unusually low C/H ratio. 

A key predictor of a Hycean atmosphere is the NH$_3$ abundance. Our results confirm previous findings\citep{Hu2021,Tsai2021} that NH$_3$ is expected to be significantly underabundant in a Hycean atmosphere due to both its photochemical destruction in a shallow atmosphere and its high solubility in the ocean underneath. NH$_3$ is also expected to be significantly underabundant compared to CH$_4$, making a high CH$_4$/NH$_3$ ratio a useful diagnostic of a Hycean atmosphere. Conversely, a significant abundance of NH$_3$ in a Hycean candidate would need to be investigated for its potential biological origins\citep{Seager2013} which are not accounted for in our models. In the absence of such a scenario, a low CH$_4$/NH$_3$ ratio may be indicative of a deep atmosphere with a low C/N ratio. 

A significant abundance of CO$_2$ is another strong predictor of a Hycean atmosphere, consistent with previous studies\citep{Hu2021}. However, we argue that the underabundance of CO$_2$ does not robustly rule out a Hycean candidate, as it can occur due to various factors, including a cold and dry atmosphere, low C content in the interior, high dissolution in the ocean and/or consumption by marine biota\citep{Bains2014,Berghuis2019}. The lack of CO$_2$, in the presence of CH$_4$ and H$_2$O would also be consistent with a deep atmosphere with a low metallicity. 

Finally, while the CO abundance on its own is not a critical diagnostic, the CO$_2$/CO ratio is expected to be higher for an Hycean atmosphere compared to that expected in the absence of an ocean (see also \citep{Hu2021}). Conversely, a CO/CO$_2$ ratio above unity would indicate a deep atmosphere with a high metallicity and high internal temperature as suggested previously\citep{Hu2021}.

In summary, considering the major molecules, the key diagnostics of a candidate Hycean world would be (a) abundance of H$_2$O, CH$_4$, other hydrocarbons and/or CO$_2$, (b) underabundance of NH$_3$, particularly a high CH$_4$/NH$_3$ ratio, and (c) CO$_2$/CO > 1. Other molecules can also be abundant depending on the specific atmospheric conditions. HCN and CH$_3$OH are expected to be underabundant in Hycean atmospheres, as also found in previous studies\citep[e.g.]{Tsai2021}. In particular, HCN can be underabundant due to the absence of NH$_3$ and CH$_3$OH can be depleted due to dissolution in the ocean. We also find that other hydrocarbons such as C$_2$H$_6$ can be abundant for cool Hycean worlds where H$_2$O may be moderately depleted in the observable atmosphere through condensation; in this case CH$_4$ photodestruction leads primarily to other hydrocarbons rather than CO and CO$_2$. 

Therefore, robustly establishing the presence of an ocean on a candidate Hycean world would require precise abundance estimates of multiple molecules discussed above and using their relative abundances to systematically constrain the key chemical pathways given the bulk properties of the planet in question. Previous work has also suggested other organic molecules as potentially observable biomarkers in Hycean worlds\citep{Madhusudhan2021}. These include molecules such as Methyl Chloride, Dimethyl Sulphide, Carbonyl Sulphide, Nitrous Oxide or Carbon Sulphide, which are produced in relatively low quantities by life on Earth but can be promising biomarkers in H$_2$-rich atmospheres, depending on the biomass present and UV enviroment\citep{Seager2013}.

\section{Summary and Discussion}
\label{sec:summary}

 Hycean worlds open a new paradigm in the search for life on exoplanets. Hycean planets are expected to be more detectable in transits and more amenable to atmospheric characterisation using transit spectroscopy than habitable rocky exoplanets. Previous work has shown that the thermodynamic conditions in the atmospheres and oceans of Hycean worlds can be similar to those known to be capable of hosting life in Earth's oceans\citep{Madhusudhan2021}. In the present work, we investigate the chemical conditions possible on a canonical Hycean world, focusing on the present and prebiotic molecular inventory in the atmosphere and the availability of bioessential elements in the ocean. The observable atmospheric composition of a Hycean planet depends on a number of factors, including the temperature structure, atmospheric metallicity, ocean surface pressure, and ocean-atmosphere interactions. However, based on the results of  several recent works\cite{Madhusudhan2021,Yu2021,Hu2021,Tsai2021} and the present study, the atmospheric compositions of Hycean worlds may be expected to include a range of prominent molecules that are detectable with modest investment of JWST time. In particular, the relative proportions of some of the key molecules, e.g. a high  CH$_4$/NH$_3$ and/or a high CO$_2$/CO, could help distinguish a Hycean world from a non-habitable mini-Neptune or super-Earth with a thicker H$_2$-rich atmosphere. 
  
\subsection{Chemical Inventory for Life}
 
We also investigate the question of whether a Hycean world would be conducive for the origin and sustenance of life. A number of factors affect the habitability of a planet, including its geophysical conditions, evolutionary history, architecture of the planetary system and astrophysical conditions \citep{Meadows2018}. From an ecological perspective, it is commonly agreed that life as we know on Earth has three essential requirements \citep{McKay2014,Cockell2016}: (a) energy sources, radiative and/or chemical (b) liquid water, and (c) various bio-essential elements. In addition to these requirements, life also needs the right environmental conditions, e.g. temperature, pressure, pH, and UV radiation. 

At the outset, by definition, the thermodynamic conditions (pressure and temperature) in a Hycean ocean would not be dissimilar to those experienced by microbial life in Earth's oceans\citep{Madhusudhan2021}. The requirements of an energy source and presence of liquid water are also naturally met, more so than most solar system environments for which a liquid water surface is a key limitation\citep{Cockell2016, Hendrix2019}. The energy source includes both the radiant energy from the star as well as the chemical free energy from chemical reactions, e.g. between abundant H$_2$ and CO$_2$ in the ocean. In particular, as discussed above, the presence of abundant H$_2$ may be more conducive to processes such as methanogenesis that are prevalent in microbial communities in anoxic environments on Earth\citep{Bains2014,Berghuis2019}. Previous studies have also discussed the feasibility of life in H$_2$-rich environments, as also seen on Earth \citep{Seager2013,Seager2020}.

Considering atmospheric evolution starting from reduced primordial conditions the atmosphere undergoes a phase rich in organic molecules potentially conducive for prebiotic chemistry. In particular, several hydrocarbons and nitriles including C$_2$H$_2$, C$_2$H$_6$, C$_3$H$_8$, C$_6$H$_6$, HCN, and HC$_3$N become abundant (above $\sim$1 ppm) starting between 1-100 Myr with some remaining abundant for over a Gyr. We note however that the abundances of these molecules are strongly dependent on the initial conditions and boundary conditions describing the ocean-atmosphere interactions. Future studies can investigate the timescales on which the primordial ocean can equilibrate with a H$_2$-rich atmosphere \citep{Kite2020,Misener2023} and how it would influence the primordial composition and chemical evolution of the atmosphere. 

We investigated the availability of bioessential nutrients in a canonical Hycean ocean building on elemental estimates for oceans of the early Earth\citep{Saito2003,Robbins2016}. Life on Earth requires a range of bioessential elements, including key volatile elements (C,H,N,O,P,S) and heavier metallic elements. While C, H, N, O and some heavier elements such as Na and K are naturally expected to be abundant in H$_2$-rich atmospheres \citep{Lodders2002,Moses2013,Madhu2016,Seager2021}, the availability of P, S, and refractory metals is a bigger challenge. In particular, for ocean worlds, the presence of a thick icy mantle below a planet-wide ocean hinders ready access to weathering of rocky surfaces that is the dominant source of bioessential metals for life in Earth's oceans\citep{Lammer2009, Maruyama2013, Noack2016, Seager2021}.

Based on internal structure models for our canonical Hycean planet, we estimate the ocean depths to be range between 100-400 km, consistent with previous studies for similar planets\citep{Noack2016,Nixon2021}. Assuming concentrations similar to those estimated for the early Earth's oceans\citep{Saito2003,Robbins2016} we estimate the mass requirements of P, S, and key bioessential refractory metals: Fe, Ni, Mn, Co, Mo, Zn and Cu. We find that the requirements can be reasonably met for plausible assumptions about the delivery of nutrients through asteroid impacts similar to the impact history of the Hadean Earth, considering a representative asteroid composition based on carbonaceous chondrites. We also consider nutrient availability in the planetary atmosphere/envelope accreted during formation assuming  solar composition, consistent with the typical $\sim$1\% dust mass fraction in protoplanetary and interstellar environments. The metals are present in condensates which  sediment out of the relatively cold Hycean atmosphere post-formation, and can meet the abundance requirement of the Hycean ocean. Making conservative assumptions about the sequestration of P and S species in the Hycean ocean, we find that atmospheric species may also make a substantial contribution to these elemental budgets. Finally, extraterrestrial dust and potential convective transport of core material through the ice layer, may also provide supplemental steady-state sources of nutrients though unlikely to meet the entire primordial  requirement. Overall, our results suggest that Hycean worlds provide adequate avenues to meet the chemical requirements of potential life similar to those estimated for the early Earth. 

\subsection{Future Directions}

Our initial investigation suggests that Hycean worlds have the potential to provide a promising environment for the origin and prevalence of life. However, the novelty of these worlds also open a wide range of open questions which affect their potential habitability. Here we outline some key directions for future research in this direction. 

\begin{itemize}

\item What are the atmospheric conditions at formation? In particular, a key question is about the timescale over which the primordial Hycean atmosphere equilibrates with the ocean underneath and what the resultant atmospheric composition could be. This sets the initial conditions for the later molecular evolution of the Hycean atmosphere, including the potential for prebiotic molecular chemistry. 

\item What are the boundary conditions at the ocean-atmosphere interface of a Hycean world? The presence of life in a Hycean ocean could add sources and sinks of prominent molecules which can significantly impact the observable chemistry. For example, methanogenesis as prevalent in anoxic environments on Earth can be a significant sink for CO$_2$ and source for additional CH$_4$ in a Hycean atmosphere. 

\item What is the possible impact history for a Hycean planet and how does the planetary system architecture influence it? Higher / lower impact fluxes or a different impactor size distribution would significantly change the bioessential element contributions. Furthermore, what happens when large bodies (10s or 100s km radius) land in very deep oceans? What percentage of the impact mass ends up in the hydrosphere?

\item Can convection of the core material through the icy mantle in a Hycean interior contribute to the bioessential element budget in the ocean? The amount of material transport through an ice VII layer is uncertain, let alone through ice X. The element flux will depend on many other factors such as the heat flux from the core, the size of the ice layer, the presence of liquid water at the bottom of the ice layer, and a host of poorly known chemical and physical properties of high-pressure ices.

\item How do (bio)(geo)chemical cycles operate on Hycean worlds? The lack of rocky surfaces makes an earth-like cycle impossible. There could be some kind of equilibrium between precipitation onto the seafloor and these precipitates having their minerals leached back into the ocean.

\item How do dynamical processes affect Hycean worlds? Accurate modeling of the atmospheric circulation is needed to properly establish the climates of Hycean planets, which in turn would have effects on the chemistry, ocean-atmosphere interactions and observable features of the atmospheres. Similarly, modeling of ocean dynamics is also needed to assess the homogeneity and turbulence of the ocean, which can have significant effects on nutrient profiles, sedimentation, and leaching rates, which in turn can affect biochemical processes.

\end{itemize}

Overall, in this work we find that the chemical conditions in Hycean atmospheres and nutrient availability in their oceans have the potential to meet the requirements for oceanic microbial life similar to those in the early Earth. As discussed above, several important questions remain and motivate new directions for further theoretical investigations on the potential habitability of Hycean worlds. Observationally, Hycean worlds show significant promise to expand and accelerate the search for signatures of extraterrestrial life with current and upcoming facilities within this decade. 

\section*{Author Contributions}
NM conceived, planned and led the project and writing of the manuscript with contributions from JM, FR and EB. NM conducted the modelling of the atmospheric structure and JM conducted the photochemical modelling with inputs from NM. JM, NM and FR processed the outputs of the photochemical calculations. FR and NM conducted the interior modelling. EB and NM conducted the estimations of bioessential elements. NM and FR computed the atmospheric spectra and synthetic JWST observations.


\section*{Conflicts of interest}
There are no conflicts to declare.


\section*{Acknowledgements}
 NM, FR and EB acknowledge support from STFC and UKRI towards the PhD studies of FR and EB. JM acknowledges support from the NASA Exoplanet Research Program grant 80NSSC23K0281. NM thanks Alexandra Turchyn, Mark Wyatt, Siddharth Gandhi, Anjali Piette, and Andrew Sellek for helpful discussions. EB thanks Skyla White and Richard Anslow for helpful discussions.



\balance


\providecommand*{\mcitethebibliography}{\thebibliography}
\csname @ifundefined\endcsname{endmcitethebibliography}
{\let\endmcitethebibliography\endthebibliography}{}


\begin{mcitethebibliography}{141}
\providecommand*{\natexlab}[1]{#1}
\providecommand*{\mciteSetBstSublistMode}[1]{}
\providecommand*{\mciteSetBstMaxWidthForm}[2]{}
\providecommand*{\mciteBstWouldAddEndPuncttrue}
  {\def\EndOfBibitem{\unskip.}}
\providecommand*{\mciteBstWouldAddEndPunctfalse}
  {\let\EndOfBibitem\relax}
\providecommand*{\mciteSetBstMidEndSepPunct}[3]{}
\providecommand*{\mciteSetBstSublistLabelBeginEnd}[3]{}
\providecommand*{\EndOfBibitem}{}
\mciteSetBstSublistMode{f}
\mciteSetBstMaxWidthForm{subitem}
{(\emph{\alph{mcitesubitemcount}})}
\mciteSetBstSublistLabelBeginEnd{\mcitemaxwidthsubitemform\space}
{\relax}{\relax}

\bibitem[Arnold \emph{et~al.}(2014)Arnold\emph{et~al.}]{Arnold2014}
L.~Arnold \emph{et~al.}, \emph{Astron. Astrophys.}, 2014, \textbf{564},
  A58\relax
\mciteBstWouldAddEndPuncttrue
\mciteSetBstMidEndSepPunct{\mcitedefaultmidpunct}
{\mcitedefaultendpunct}{\mcitedefaultseppunct}\relax
\EndOfBibitem
\bibitem[Rodler and L{\'o}pez-Morales(2014)]{Rodler2014}
F.~Rodler and M.~L{\'o}pez-Morales, \emph{Astrophys. J.}, 2014, \textbf{781},
  54\relax
\mciteBstWouldAddEndPuncttrue
\mciteSetBstMidEndSepPunct{\mcitedefaultmidpunct}
{\mcitedefaultendpunct}{\mcitedefaultseppunct}\relax
\EndOfBibitem
\bibitem[Feng \emph{et~al.}(2018)Feng\emph{et~al.}]{Feng2018}
Y.~K. Feng \emph{et~al.}, \emph{Astron. J.}, 2018, \textbf{155}, 200\relax
\mciteBstWouldAddEndPuncttrue
\mciteSetBstMidEndSepPunct{\mcitedefaultmidpunct}
{\mcitedefaultendpunct}{\mcitedefaultseppunct}\relax
\EndOfBibitem
\bibitem[{Kasting} \emph{et~al.}(1993){Kasting}, {Whitmire}, and
  {Reynolds}]{Kasting1993}
J.~F. {Kasting}, D.~P. {Whitmire} and R.~T. {Reynolds}, \emph{Icarus}, 1993,
  \textbf{101}, 108--128\relax
\mciteBstWouldAddEndPuncttrue
\mciteSetBstMidEndSepPunct{\mcitedefaultmidpunct}
{\mcitedefaultendpunct}{\mcitedefaultseppunct}\relax
\EndOfBibitem
\bibitem[{Selsis} \emph{et~al.}(2007){Selsis}\emph{et~al.}]{selsis2007}
F.~{Selsis} \emph{et~al.}, \emph{Astron. Astrophys.}, 2007, \textbf{476},
  1373--1387\relax
\mciteBstWouldAddEndPuncttrue
\mciteSetBstMidEndSepPunct{\mcitedefaultmidpunct}
{\mcitedefaultendpunct}{\mcitedefaultseppunct}\relax
\EndOfBibitem
\bibitem[{Kopparapu}
  \emph{et~al.}(2013){Kopparapu}\emph{et~al.}]{Kopparapu2013}
R.~K. {Kopparapu} \emph{et~al.}, \emph{Astrophys. J.}, 2013, \textbf{765},
  131\relax
\mciteBstWouldAddEndPuncttrue
\mciteSetBstMidEndSepPunct{\mcitedefaultmidpunct}
{\mcitedefaultendpunct}{\mcitedefaultseppunct}\relax
\EndOfBibitem
\bibitem[{Elkins-Tanton} and {Seager}(2008)]{elkins-tanton2008}
L.~T. {Elkins-Tanton} and S.~{Seager}, \emph{Astrophys. J.}, 2008,
  \textbf{685}, 1237--1246\relax
\mciteBstWouldAddEndPuncttrue
\mciteSetBstMidEndSepPunct{\mcitedefaultmidpunct}
{\mcitedefaultendpunct}{\mcitedefaultseppunct}\relax
\EndOfBibitem
\bibitem[Gillon \emph{et~al.}(2017)Gillon\emph{et~al.}]{Gillon2017}
M.~Gillon \emph{et~al.}, \emph{Nature}, 2017, \textbf{542}, 456--460\relax
\mciteBstWouldAddEndPuncttrue
\mciteSetBstMidEndSepPunct{\mcitedefaultmidpunct}
{\mcitedefaultendpunct}{\mcitedefaultseppunct}\relax
\EndOfBibitem
\bibitem[Dittmann \emph{et~al.}(2017)Dittmann\emph{et~al.}]{Dittmann2017}
J.~A. Dittmann \emph{et~al.}, \emph{Nature}, 2017, \textbf{544}, 333--336\relax
\mciteBstWouldAddEndPuncttrue
\mciteSetBstMidEndSepPunct{\mcitedefaultmidpunct}
{\mcitedefaultendpunct}{\mcitedefaultseppunct}\relax
\EndOfBibitem
\bibitem[Ment \emph{et~al.}(2019)Ment\emph{et~al.}]{Ment2019}
K.~Ment \emph{et~al.}, \emph{Astron. J.}, 2019, \textbf{157}, 32\relax
\mciteBstWouldAddEndPuncttrue
\mciteSetBstMidEndSepPunct{\mcitedefaultmidpunct}
{\mcitedefaultendpunct}{\mcitedefaultseppunct}\relax
\EndOfBibitem
\bibitem[Lillo-Box \emph{et~al.}(2020)Lillo-Box\emph{et~al.}]{Lillo2020}
J.~Lillo-Box \emph{et~al.}, \emph{Astron. Astrophys.}, 2020, \textbf{642},
  A121\relax
\mciteBstWouldAddEndPuncttrue
\mciteSetBstMidEndSepPunct{\mcitedefaultmidpunct}
{\mcitedefaultendpunct}{\mcitedefaultseppunct}\relax
\EndOfBibitem
\bibitem[Gilbert \emph{et~al.}(2020)Gilbert\emph{et~al.}]{Gilbert2020}
E.~A. Gilbert \emph{et~al.}, \emph{Astron. J.}, 2020, \textbf{160}, 116\relax
\mciteBstWouldAddEndPuncttrue
\mciteSetBstMidEndSepPunct{\mcitedefaultmidpunct}
{\mcitedefaultendpunct}{\mcitedefaultseppunct}\relax
\EndOfBibitem
\bibitem[Rodriguez \emph{et~al.}(2020)Rodriguez\emph{et~al.}]{Rodriguez2020}
J.~E. Rodriguez \emph{et~al.}, \emph{Astron. J.}, 2020, \textbf{160}, 117\relax
\mciteBstWouldAddEndPuncttrue
\mciteSetBstMidEndSepPunct{\mcitedefaultmidpunct}
{\mcitedefaultendpunct}{\mcitedefaultseppunct}\relax
\EndOfBibitem
\bibitem[Barstow and Irwin(2016)]{Barstow2016}
J.~K. Barstow and P.~G. Irwin, \emph{Mon. Notices Royal Astron. Soc.}, 2016,
  \textbf{461}, L92--L96\relax
\mciteBstWouldAddEndPuncttrue
\mciteSetBstMidEndSepPunct{\mcitedefaultmidpunct}
{\mcitedefaultendpunct}{\mcitedefaultseppunct}\relax
\EndOfBibitem
\bibitem[{Snellen} \emph{et~al.}(2017){Snellen}\emph{et~al.}]{Snellen2017}
I.~A.~G. {Snellen} \emph{et~al.}, \emph{Astron. J.}, 2017, \textbf{154},
  77\relax
\mciteBstWouldAddEndPuncttrue
\mciteSetBstMidEndSepPunct{\mcitedefaultmidpunct}
{\mcitedefaultendpunct}{\mcitedefaultseppunct}\relax
\EndOfBibitem
\bibitem[Lustig-Yaeger
  \emph{et~al.}(2019)Lustig-Yaeger\emph{et~al.}]{Lustig2019}
J.~Lustig-Yaeger \emph{et~al.}, \emph{Astron. J.}, 2019, \textbf{158}, 27\relax
\mciteBstWouldAddEndPuncttrue
\mciteSetBstMidEndSepPunct{\mcitedefaultmidpunct}
{\mcitedefaultendpunct}{\mcitedefaultseppunct}\relax
\EndOfBibitem
\bibitem[{Madhusudhan} \emph{et~al.}(2021){Madhusudhan}, {Piette}, and
  {Constantinou}]{Madhusudhan2021}
N.~{Madhusudhan}, A.~A.~A. {Piette} and S.~{Constantinou}, \emph{Astrophys.
  J.}, 2021, \textbf{918}, 1\relax
\mciteBstWouldAddEndPuncttrue
\mciteSetBstMidEndSepPunct{\mcitedefaultmidpunct}
{\mcitedefaultendpunct}{\mcitedefaultseppunct}\relax
\EndOfBibitem
\bibitem[{Madhusudhan}
  \emph{et~al.}(2020){Madhusudhan}\emph{et~al.}]{Madhusudhan2020}
N.~{Madhusudhan} \emph{et~al.}, \emph{Astrophys. J.}, 2020, \textbf{891},
  L7\relax
\mciteBstWouldAddEndPuncttrue
\mciteSetBstMidEndSepPunct{\mcitedefaultmidpunct}
{\mcitedefaultendpunct}{\mcitedefaultseppunct}\relax
\EndOfBibitem
\bibitem[{Piette} and {Madhusudhan}(2020)]{Piette2020}
A.~A.~A. {Piette} and N.~{Madhusudhan}, \emph{Astrophys. J.}, 2020,
  \textbf{904}, 154\relax
\mciteBstWouldAddEndPuncttrue
\mciteSetBstMidEndSepPunct{\mcitedefaultmidpunct}
{\mcitedefaultendpunct}{\mcitedefaultseppunct}\relax
\EndOfBibitem
\bibitem[{Nixon} and {Madhusudhan}(2021)]{Nixon2021}
M.~C. {Nixon} and N.~{Madhusudhan}, \emph{Mon. Notices Royal Astron. Soc.},
  2021, \textbf{505}, 3414--3432\relax
\mciteBstWouldAddEndPuncttrue
\mciteSetBstMidEndSepPunct{\mcitedefaultmidpunct}
{\mcitedefaultendpunct}{\mcitedefaultseppunct}\relax
\EndOfBibitem
\bibitem[{Stevenson}(1999)]{stevenson1999}
D.~J. {Stevenson}, \emph{Nature}, 1999, \textbf{400}, 32\relax
\mciteBstWouldAddEndPuncttrue
\mciteSetBstMidEndSepPunct{\mcitedefaultmidpunct}
{\mcitedefaultendpunct}{\mcitedefaultseppunct}\relax
\EndOfBibitem
\bibitem[{Pierrehumbert} and {Gaidos}(2011)]{pierrehumbert2011}
R.~{Pierrehumbert} and E.~{Gaidos}, \emph{Astrophys. J. Lett.}, 2011,
  \textbf{734}, L13\relax
\mciteBstWouldAddEndPuncttrue
\mciteSetBstMidEndSepPunct{\mcitedefaultmidpunct}
{\mcitedefaultendpunct}{\mcitedefaultseppunct}\relax
\EndOfBibitem
\bibitem[{Mol Lous} \emph{et~al.}(2022){Mol Lous}, {Helled}, and
  {Mordasini}]{Mol_Lous2022}
M.~{Mol Lous}, R.~{Helled} and C.~{Mordasini}, \emph{Nat. Astron.}, 2022,
  \textbf{6}, 819--827\relax
\mciteBstWouldAddEndPuncttrue
\mciteSetBstMidEndSepPunct{\mcitedefaultmidpunct}
{\mcitedefaultendpunct}{\mcitedefaultseppunct}\relax
\EndOfBibitem
\bibitem[{Fressin} \emph{et~al.}(2013){Fressin}\emph{et~al.}]{Fressin2013}
F.~{Fressin} \emph{et~al.}, \emph{Astrophys. J.}, 2013, \textbf{766}, 81\relax
\mciteBstWouldAddEndPuncttrue
\mciteSetBstMidEndSepPunct{\mcitedefaultmidpunct}
{\mcitedefaultendpunct}{\mcitedefaultseppunct}\relax
\EndOfBibitem
\bibitem[{Fulton} and {Petigura}(2018)]{fulton2018}
B.~J. {Fulton} and E.~A. {Petigura}, \emph{AJ}, 2018, \textbf{156}, 264\relax
\mciteBstWouldAddEndPuncttrue
\mciteSetBstMidEndSepPunct{\mcitedefaultmidpunct}
{\mcitedefaultendpunct}{\mcitedefaultseppunct}\relax
\EndOfBibitem
\bibitem[{Borucki} \emph{et~al.}(2010){Borucki}\emph{et~al.}]{Borucki2010}
W.~J. {Borucki} \emph{et~al.}, \emph{Science}, 2010, \textbf{327}, 977\relax
\mciteBstWouldAddEndPuncttrue
\mciteSetBstMidEndSepPunct{\mcitedefaultmidpunct}
{\mcitedefaultendpunct}{\mcitedefaultseppunct}\relax
\EndOfBibitem
\bibitem[{Ricker} \emph{et~al.}(2015){Ricker}\emph{et~al.}]{Ricker2015}
G.~R. {Ricker} \emph{et~al.}, \emph{J. Astron. Telesc. Instrum.Syst.}, 2015,
  \textbf{1}, 014003\relax
\mciteBstWouldAddEndPuncttrue
\mciteSetBstMidEndSepPunct{\mcitedefaultmidpunct}
{\mcitedefaultendpunct}{\mcitedefaultseppunct}\relax
\EndOfBibitem
\bibitem[{Fukui} \emph{et~al.}(2022){Fukui}\emph{et~al.}]{Fukui2022}
A.~{Fukui} \emph{et~al.}, \emph{Publ. Astron. Soc. Japan}, 2022, \textbf{74},
  L1--L8\relax
\mciteBstWouldAddEndPuncttrue
\mciteSetBstMidEndSepPunct{\mcitedefaultmidpunct}
{\mcitedefaultendpunct}{\mcitedefaultseppunct}\relax
\EndOfBibitem
\bibitem[{Kawauchi} \emph{et~al.}(2022){Kawauchi}\emph{et~al.}]{Kawauchi2022}
K.~{Kawauchi} \emph{et~al.}, \emph{Astron. Astrophys.}, 2022, \textbf{666},
  A4\relax
\mciteBstWouldAddEndPuncttrue
\mciteSetBstMidEndSepPunct{\mcitedefaultmidpunct}
{\mcitedefaultendpunct}{\mcitedefaultseppunct}\relax
\EndOfBibitem
\bibitem[{Mikal-Evans}
  \emph{et~al.}(2023){Mikal-Evans}\emph{et~al.}]{Evans2023}
T.~{Mikal-Evans} \emph{et~al.}, \emph{Astron. J.}, 2023, \textbf{165}, 84\relax
\mciteBstWouldAddEndPuncttrue
\mciteSetBstMidEndSepPunct{\mcitedefaultmidpunct}
{\mcitedefaultendpunct}{\mcitedefaultseppunct}\relax
\EndOfBibitem
\bibitem[{Piaulet} \emph{et~al.}(2023){Piaulet}\emph{et~al.}]{Piaulet2023}
C.~{Piaulet} \emph{et~al.}, \emph{Nat. Astron.}, 2023, \textbf{7},
  206--222\relax
\mciteBstWouldAddEndPuncttrue
\mciteSetBstMidEndSepPunct{\mcitedefaultmidpunct}
{\mcitedefaultendpunct}{\mcitedefaultseppunct}\relax
\EndOfBibitem
\bibitem[{Phillips} \emph{et~al.}(2021){Phillips}\emph{et~al.}]{Phillips2021}
C.~L. {Phillips} \emph{et~al.}, \emph{Astrophys. J.}, 2021, \textbf{923},
  144\relax
\mciteBstWouldAddEndPuncttrue
\mciteSetBstMidEndSepPunct{\mcitedefaultmidpunct}
{\mcitedefaultendpunct}{\mcitedefaultseppunct}\relax
\EndOfBibitem
\bibitem[{Phillips} \emph{et~al.}(2022){Phillips}\emph{et~al.}]{Phillips2022}
C.~{Phillips} \emph{et~al.}, \emph{arXiv e-prints}, 2022,
  arXiv:2209.12919\relax
\mciteBstWouldAddEndPuncttrue
\mciteSetBstMidEndSepPunct{\mcitedefaultmidpunct}
{\mcitedefaultendpunct}{\mcitedefaultseppunct}\relax
\EndOfBibitem
\bibitem[{Leung} \emph{et~al.}(2022){Leung}\emph{et~al.}]{Leung2022}
M.~{Leung} \emph{et~al.}, \emph{Astrophys. J.}, 2022, \textbf{938}, 6\relax
\mciteBstWouldAddEndPuncttrue
\mciteSetBstMidEndSepPunct{\mcitedefaultmidpunct}
{\mcitedefaultendpunct}{\mcitedefaultseppunct}\relax
\EndOfBibitem
\bibitem[{Yu} \emph{et~al.}(2021){Yu}\emph{et~al.}]{Yu2021}
X.~{Yu} \emph{et~al.}, \emph{Astrophys. J.}, 2021, \textbf{914}, 38\relax
\mciteBstWouldAddEndPuncttrue
\mciteSetBstMidEndSepPunct{\mcitedefaultmidpunct}
{\mcitedefaultendpunct}{\mcitedefaultseppunct}\relax
\EndOfBibitem
\bibitem[{Hu} \emph{et~al.}(2021){Hu}\emph{et~al.}]{Hu2021}
R.~{Hu} \emph{et~al.}, \emph{Astrophys. J. Lett.}, 2021, \textbf{921}, L8\relax
\mciteBstWouldAddEndPuncttrue
\mciteSetBstMidEndSepPunct{\mcitedefaultmidpunct}
{\mcitedefaultendpunct}{\mcitedefaultseppunct}\relax
\EndOfBibitem
\bibitem[{Tsai} \emph{et~al.}(2021){Tsai}\emph{et~al.}]{Tsai2021}
S.-M. {Tsai} \emph{et~al.}, \emph{Astrophys. J. Lett.}, 2021, \textbf{922},
  L27\relax
\mciteBstWouldAddEndPuncttrue
\mciteSetBstMidEndSepPunct{\mcitedefaultmidpunct}
{\mcitedefaultendpunct}{\mcitedefaultseppunct}\relax
\EndOfBibitem
\bibitem[{McKay}(2014)]{McKay2014}
C.~P. {McKay}, \emph{Proc. Natl. Acad. Sci.}, 2014, \textbf{111},
  12628--12633\relax
\mciteBstWouldAddEndPuncttrue
\mciteSetBstMidEndSepPunct{\mcitedefaultmidpunct}
{\mcitedefaultendpunct}{\mcitedefaultseppunct}\relax
\EndOfBibitem
\bibitem[{Cockell} \emph{et~al.}(2016){Cockell}\emph{et~al.}]{Cockell2016}
C.~S. {Cockell} \emph{et~al.}, \emph{Astrobiology}, 2016, \textbf{16},
  89--117\relax
\mciteBstWouldAddEndPuncttrue
\mciteSetBstMidEndSepPunct{\mcitedefaultmidpunct}
{\mcitedefaultendpunct}{\mcitedefaultseppunct}\relax
\EndOfBibitem
\bibitem[{Hendrix} \emph{et~al.}(2019){Hendrix}\emph{et~al.}]{Hendrix2019}
A.~R. {Hendrix} \emph{et~al.}, \emph{Astrobiology}, 2019, \textbf{19},
  1--27\relax
\mciteBstWouldAddEndPuncttrue
\mciteSetBstMidEndSepPunct{\mcitedefaultmidpunct}
{\mcitedefaultendpunct}{\mcitedefaultseppunct}\relax
\EndOfBibitem
\bibitem[Lammer \emph{et~al.}(2009)Lammer\emph{et~al.}]{Lammer2009}
H.~Lammer \emph{et~al.}, \emph{Astron Astrophys Rev}, 2009, \textbf{17},
  181--249\relax
\mciteBstWouldAddEndPuncttrue
\mciteSetBstMidEndSepPunct{\mcitedefaultmidpunct}
{\mcitedefaultendpunct}{\mcitedefaultseppunct}\relax
\EndOfBibitem
\bibitem[{Noack} \emph{et~al.}(2016){Noack}\emph{et~al.}]{Noack2016}
L.~{Noack} \emph{et~al.}, \emph{Icarus}, 2016, \textbf{277}, 215--236\relax
\mciteBstWouldAddEndPuncttrue
\mciteSetBstMidEndSepPunct{\mcitedefaultmidpunct}
{\mcitedefaultendpunct}{\mcitedefaultseppunct}\relax
\EndOfBibitem
\bibitem[{Lingam} and {Loeb}(2018)]{Lingam2018}
M.~{Lingam} and A.~{Loeb}, \emph{Astron. J.}, 2018, \textbf{156}, 151\relax
\mciteBstWouldAddEndPuncttrue
\mciteSetBstMidEndSepPunct{\mcitedefaultmidpunct}
{\mcitedefaultendpunct}{\mcitedefaultseppunct}\relax
\EndOfBibitem
\bibitem[{Journaux} \emph{et~al.}(2020){Journaux}\emph{et~al.}]{Journaux2020}
B.~{Journaux} \emph{et~al.}, \emph{Space Sci. Rev.}, 2020, \textbf{216},
  7\relax
\mciteBstWouldAddEndPuncttrue
\mciteSetBstMidEndSepPunct{\mcitedefaultmidpunct}
{\mcitedefaultendpunct}{\mcitedefaultseppunct}\relax
\EndOfBibitem
\bibitem[Maruyama \emph{et~al.}(2013)Maruyama\emph{et~al.}]{Maruyama2013}
S.~Maruyama \emph{et~al.}, \emph{Geosci. Front.}, 2013, \textbf{4},
  141--165\relax
\mciteBstWouldAddEndPuncttrue
\mciteSetBstMidEndSepPunct{\mcitedefaultmidpunct}
{\mcitedefaultendpunct}{\mcitedefaultseppunct}\relax
\EndOfBibitem
\bibitem[{Seager} \emph{et~al.}(2021){Seager}\emph{et~al.}]{Seager2021}
S.~{Seager} \emph{et~al.}, \emph{Universe}, 2021, \textbf{7}, 172\relax
\mciteBstWouldAddEndPuncttrue
\mciteSetBstMidEndSepPunct{\mcitedefaultmidpunct}
{\mcitedefaultendpunct}{\mcitedefaultseppunct}\relax
\EndOfBibitem
\bibitem[Choblet \emph{et~al.}(2017)Choblet\emph{et~al.}]{Choblet2017}
G.~Choblet \emph{et~al.}, \emph{Icarus}, 2017, \textbf{285}, 252--262\relax
\mciteBstWouldAddEndPuncttrue
\mciteSetBstMidEndSepPunct{\mcitedefaultmidpunct}
{\mcitedefaultendpunct}{\mcitedefaultseppunct}\relax
\EndOfBibitem
\bibitem[Kalousová \emph{et~al.}(2018)Kalousová\emph{et~al.}]{Kalousova2018}
K.~Kalousová \emph{et~al.}, \emph{Icarus}, 2018, \textbf{299}, 133--147\relax
\mciteBstWouldAddEndPuncttrue
\mciteSetBstMidEndSepPunct{\mcitedefaultmidpunct}
{\mcitedefaultendpunct}{\mcitedefaultseppunct}\relax
\EndOfBibitem
\bibitem[Kalousová and Sotin(2018)]{KS2018}
K.~Kalousová and C.~Sotin, \emph{Geophys. Res. Lett.}, 2018, \textbf{45},
  8096--8103\relax
\mciteBstWouldAddEndPuncttrue
\mciteSetBstMidEndSepPunct{\mcitedefaultmidpunct}
{\mcitedefaultendpunct}{\mcitedefaultseppunct}\relax
\EndOfBibitem
\bibitem[{Hernandez} \emph{et~al.}(2022){Hernandez}, {Caracas}, and
  {Labrosse}]{Hernandez2022}
J.-A. {Hernandez}, R.~{Caracas} and S.~{Labrosse}, \emph{Nature
  Communications}, 2022, \textbf{13}, 3303\relax
\mciteBstWouldAddEndPuncttrue
\mciteSetBstMidEndSepPunct{\mcitedefaultmidpunct}
{\mcitedefaultendpunct}{\mcitedefaultseppunct}\relax
\EndOfBibitem
\bibitem[{Lebec} \emph{et~al.}(2023){Lebec}\emph{et~al.}]{Lebec2023}
L.~{Lebec} \emph{et~al.}, \emph{Icarus}, 2023, \textbf{396}, 115494\relax
\mciteBstWouldAddEndPuncttrue
\mciteSetBstMidEndSepPunct{\mcitedefaultmidpunct}
{\mcitedefaultendpunct}{\mcitedefaultseppunct}\relax
\EndOfBibitem
\bibitem[{Rimmer} \emph{et~al.}(2018){Rimmer}\emph{et~al.}]{Rimmer2018}
P.~B. {Rimmer} \emph{et~al.}, \emph{Sci. Adv.}, 2018, \textbf{4},
  eaar3302\relax
\mciteBstWouldAddEndPuncttrue
\mciteSetBstMidEndSepPunct{\mcitedefaultmidpunct}
{\mcitedefaultendpunct}{\mcitedefaultseppunct}\relax
\EndOfBibitem
\bibitem[{Parkos} \emph{et~al.}(2018){Parkos}\emph{et~al.}]{Parkos2018}
D.~{Parkos} \emph{et~al.}, \emph{Journal of Geophysical Research (Planets)},
  2018, \textbf{123}, 892--909\relax
\mciteBstWouldAddEndPuncttrue
\mciteSetBstMidEndSepPunct{\mcitedefaultmidpunct}
{\mcitedefaultendpunct}{\mcitedefaultseppunct}\relax
\EndOfBibitem
\bibitem[{Innes} \emph{et~al.}(2023){Innes}, {Tsai}, and
  {Pierrehumbert}]{Innes2023}
H.~{Innes}, S.-M. {Tsai} and R.~T. {Pierrehumbert}, \emph{arXiv e-prints},
  2023,  arXiv:2304.02698\relax
\mciteBstWouldAddEndPuncttrue
\mciteSetBstMidEndSepPunct{\mcitedefaultmidpunct}
{\mcitedefaultendpunct}{\mcitedefaultseppunct}\relax
\EndOfBibitem
\bibitem[{Gandhi} and {Madhusudhan}(2017)]{Gandhi2017}
S.~{Gandhi} and N.~{Madhusudhan}, \emph{Mon. Notices Royal Astron. Soc.}, 2017,
  \textbf{472}, 2334--2355\relax
\mciteBstWouldAddEndPuncttrue
\mciteSetBstMidEndSepPunct{\mcitedefaultmidpunct}
{\mcitedefaultendpunct}{\mcitedefaultseppunct}\relax
\EndOfBibitem
\bibitem[{Hubeny}(2017)]{Hubeny2017}
I.~{Hubeny}, \emph{Mon. Notices Royal Astron. Soc.}, 2017, \textbf{469},
  841--869\relax
\mciteBstWouldAddEndPuncttrue
\mciteSetBstMidEndSepPunct{\mcitedefaultmidpunct}
{\mcitedefaultendpunct}{\mcitedefaultseppunct}\relax
\EndOfBibitem
\bibitem[{Castor} \emph{et~al.}(1992){Castor}, {Dykema}, and
  {Klein}]{Castor1992}
J.~I. {Castor}, P.~G. {Dykema} and R.~I. {Klein}, \emph{Astrophys. J.}, 1992,
  \textbf{387}, 561\relax
\mciteBstWouldAddEndPuncttrue
\mciteSetBstMidEndSepPunct{\mcitedefaultmidpunct}
{\mcitedefaultendpunct}{\mcitedefaultseppunct}\relax
\EndOfBibitem
\bibitem[{Benneke} \emph{et~al.}(2019){Benneke}\emph{et~al.}]{Benneke2019}
B.~{Benneke} \emph{et~al.}, \emph{Astrophys. J. Lett.}, 2019, \textbf{887},
  L14\relax
\mciteBstWouldAddEndPuncttrue
\mciteSetBstMidEndSepPunct{\mcitedefaultmidpunct}
{\mcitedefaultendpunct}{\mcitedefaultseppunct}\relax
\EndOfBibitem
\bibitem[{Cloutier} \emph{et~al.}(2019){Cloutier}\emph{et~al.}]{Cloutier2019}
R.~{Cloutier} \emph{et~al.}, \emph{Astron. Astrophys.}, 2019, \textbf{621},
  A49\relax
\mciteBstWouldAddEndPuncttrue
\mciteSetBstMidEndSepPunct{\mcitedefaultmidpunct}
{\mcitedefaultendpunct}{\mcitedefaultseppunct}\relax
\EndOfBibitem
\bibitem[{Barber} \emph{et~al.}(2006){Barber}, {Tennyson}, {Harris}, and
  {Tolchenov}]{barber2006}
R.~J. {Barber}, J.~{Tennyson}, G.~J. {Harris} and R.~N. {Tolchenov}, \emph{Mon.
  Notices Royal Astron. Soc.}, 2006, \textbf{368}, 1087--1094\relax
\mciteBstWouldAddEndPuncttrue
\mciteSetBstMidEndSepPunct{\mcitedefaultmidpunct}
{\mcitedefaultendpunct}{\mcitedefaultseppunct}\relax
\EndOfBibitem
\bibitem[Rothman \emph{et~al.}(2010)Rothman\emph{et~al.}]{rothman_hitemp_2010}
L.~Rothman \emph{et~al.}, \emph{J. Quant. Spectrosc. Radiat. Transf.}, 2010,
  \textbf{111}, 2139--2150\relax
\mciteBstWouldAddEndPuncttrue
\mciteSetBstMidEndSepPunct{\mcitedefaultmidpunct}
{\mcitedefaultendpunct}{\mcitedefaultseppunct}\relax
\EndOfBibitem
\bibitem[{Borysow} \emph{et~al.}(1988){Borysow}, {Frommhold}, and
  {Birnbaum}]{borysow1988}
J.~{Borysow}, L.~{Frommhold} and G.~{Birnbaum}, \emph{Astrophys. J.}, 1988,
  \textbf{326}, 509\relax
\mciteBstWouldAddEndPuncttrue
\mciteSetBstMidEndSepPunct{\mcitedefaultmidpunct}
{\mcitedefaultendpunct}{\mcitedefaultseppunct}\relax
\EndOfBibitem
\bibitem[{Orton} \emph{et~al.}(2007){Orton}\emph{et~al.}]{orton2007}
G.~S. {Orton} \emph{et~al.}, \emph{Icarus}, 2007, \textbf{189}, 544--549\relax
\mciteBstWouldAddEndPuncttrue
\mciteSetBstMidEndSepPunct{\mcitedefaultmidpunct}
{\mcitedefaultendpunct}{\mcitedefaultseppunct}\relax
\EndOfBibitem
\bibitem[{Abel} \emph{et~al.}(2011){Abel}, {Frommhold}, {Li}, and
  {Hunt}]{abel2011}
M.~{Abel}, L.~{Frommhold}, X.~{Li} and K.~L.~C. {Hunt}, \emph{J. Phys. Chem.
  A}, 2011, \textbf{115}, 6805--6812\relax
\mciteBstWouldAddEndPuncttrue
\mciteSetBstMidEndSepPunct{\mcitedefaultmidpunct}
{\mcitedefaultendpunct}{\mcitedefaultseppunct}\relax
\EndOfBibitem
\bibitem[Richard and others.(2012)]{richard_new_2012}
C.~Richard and others., \emph{J. Quant. Spectrosc. Radiat. Transf.}, 2012,
  \textbf{113}, 1276--1285\relax
\mciteBstWouldAddEndPuncttrue
\mciteSetBstMidEndSepPunct{\mcitedefaultmidpunct}
{\mcitedefaultendpunct}{\mcitedefaultseppunct}\relax
\EndOfBibitem
\bibitem[{Allen} \emph{et~al.}(1981){Allen}, {Yung}, and {Waters}]{allen81}
M.~{Allen}, Y.~L. {Yung} and J.~W. {Waters}, \emph{J. Geophys. Res.}, 1981,
  \textbf{86}, 3617--3627\relax
\mciteBstWouldAddEndPuncttrue
\mciteSetBstMidEndSepPunct{\mcitedefaultmidpunct}
{\mcitedefaultendpunct}{\mcitedefaultseppunct}\relax
\EndOfBibitem
\bibitem[{Yung} \emph{et~al.}(1984){Yung}, {Allen}, and {Pinto}]{yung84}
Y.~L. {Yung}, M.~{Allen} and J.~P. {Pinto}, \emph{Astrophys. J. Suppl.}, 1984,
  \textbf{55}, 465--506\relax
\mciteBstWouldAddEndPuncttrue
\mciteSetBstMidEndSepPunct{\mcitedefaultmidpunct}
{\mcitedefaultendpunct}{\mcitedefaultseppunct}\relax
\EndOfBibitem
\bibitem[{Moses} \emph{et~al.}(2011){Moses}\emph{et~al.}]{moses11}
J.~I. {Moses} \emph{et~al.}, \emph{Astrophys. J.}, 2011, \textbf{737}, 15\relax
\mciteBstWouldAddEndPuncttrue
\mciteSetBstMidEndSepPunct{\mcitedefaultmidpunct}
{\mcitedefaultendpunct}{\mcitedefaultseppunct}\relax
\EndOfBibitem
\bibitem[{Kite} and {Ford}(2018)]{Kite2018}
E.~S. {Kite} and E.~B. {Ford}, \emph{Astrophys. J.}, 2018, \textbf{864},
  75\relax
\mciteBstWouldAddEndPuncttrue
\mciteSetBstMidEndSepPunct{\mcitedefaultmidpunct}
{\mcitedefaultendpunct}{\mcitedefaultseppunct}\relax
\EndOfBibitem
\bibitem[{Moses} \emph{et~al.}(2013){Moses}\emph{et~al.}]{moses13}
J.~I. {Moses} \emph{et~al.}, \emph{Astrophys. J.}, 2013, \textbf{777}, 34\relax
\mciteBstWouldAddEndPuncttrue
\mciteSetBstMidEndSepPunct{\mcitedefaultmidpunct}
{\mcitedefaultendpunct}{\mcitedefaultseppunct}\relax
\EndOfBibitem
\bibitem[{Kite} \emph{et~al.}(2009){Kite}, {Manga}, and {Gaidos}]{Kite2009}
E.~S. {Kite}, M.~{Manga} and E.~{Gaidos}, \emph{Astrophys. J.}, 2009,
  \textbf{700}, 1732--1749\relax
\mciteBstWouldAddEndPuncttrue
\mciteSetBstMidEndSepPunct{\mcitedefaultmidpunct}
{\mcitedefaultendpunct}{\mcitedefaultseppunct}\relax
\EndOfBibitem
\bibitem[{Moses} \emph{et~al.}(2022){Moses}\emph{et~al.}]{moses22tremgrid}
J.~I. {Moses} \emph{et~al.}, \emph{Exp. Astron.}, 2022, \textbf{53},
  279--322\relax
\mciteBstWouldAddEndPuncttrue
\mciteSetBstMidEndSepPunct{\mcitedefaultmidpunct}
{\mcitedefaultendpunct}{\mcitedefaultseppunct}\relax
\EndOfBibitem
\bibitem[{Peacock} \emph{et~al.}(2020){Peacock}\emph{et~al.}]{peacock20}
S.~{Peacock} \emph{et~al.}, \emph{Astrophys. J.}, 2020, \textbf{895}, 5\relax
\mciteBstWouldAddEndPuncttrue
\mciteSetBstMidEndSepPunct{\mcitedefaultmidpunct}
{\mcitedefaultendpunct}{\mcitedefaultseppunct}\relax
\EndOfBibitem
\bibitem[{Moses} \emph{et~al.}(2000){Moses}\emph{et~al.}]{moses00b}
J.~I. {Moses} \emph{et~al.}, \emph{Icarus}, 2000, \textbf{145}, 166--202\relax
\mciteBstWouldAddEndPuncttrue
\mciteSetBstMidEndSepPunct{\mcitedefaultmidpunct}
{\mcitedefaultendpunct}{\mcitedefaultseppunct}\relax
\EndOfBibitem
\bibitem[{Kasting}(1990)]{Kasting1990}
J.~F. {Kasting}, \emph{Orig. Life Evol. Biosph.}, 1990, \textbf{20},
  199--231\relax
\mciteBstWouldAddEndPuncttrue
\mciteSetBstMidEndSepPunct{\mcitedefaultmidpunct}
{\mcitedefaultendpunct}{\mcitedefaultseppunct}\relax
\EndOfBibitem
\bibitem[{Arney} \emph{et~al.}(2016){Arney}\emph{et~al.}]{Arney2016}
G.~{Arney} \emph{et~al.}, \emph{Astrobiology}, 2016, \textbf{16},
  873--899\relax
\mciteBstWouldAddEndPuncttrue
\mciteSetBstMidEndSepPunct{\mcitedefaultmidpunct}
{\mcitedefaultendpunct}{\mcitedefaultseppunct}\relax
\EndOfBibitem
\bibitem[{Ranjan} \emph{et~al.}(2020){Ranjan}\emph{et~al.}]{Ranjan2020}
S.~{Ranjan} \emph{et~al.}, \emph{Astrophys. J.}, 2020, \textbf{896}, 148\relax
\mciteBstWouldAddEndPuncttrue
\mciteSetBstMidEndSepPunct{\mcitedefaultmidpunct}
{\mcitedefaultendpunct}{\mcitedefaultseppunct}\relax
\EndOfBibitem
\bibitem[{Madhusudhan} \emph{et~al.}(2012){Madhusudhan}, {Lee}, and
  {Mousis}]{Madhusudhan2012}
N.~{Madhusudhan}, K.~K.~M. {Lee} and O.~{Mousis}, \emph{Astrophys. J. Lett.},
  2012, \textbf{759}, L40\relax
\mciteBstWouldAddEndPuncttrue
\mciteSetBstMidEndSepPunct{\mcitedefaultmidpunct}
{\mcitedefaultendpunct}{\mcitedefaultseppunct}\relax
\EndOfBibitem
\bibitem[{Chabrier} \emph{et~al.}(2019){Chabrier}, {Mazevet}, and
  {Soubiran}]{Chabrier2019}
G.~{Chabrier}, S.~{Mazevet} and F.~{Soubiran}, \emph{Astrophys. J.}, 2019,
  \textbf{872}, 51\relax
\mciteBstWouldAddEndPuncttrue
\mciteSetBstMidEndSepPunct{\mcitedefaultmidpunct}
{\mcitedefaultendpunct}{\mcitedefaultseppunct}\relax
\EndOfBibitem
\bibitem[{Thomas} and {Madhusudhan}(2016)]{Thomas2016}
S.~W. {Thomas} and N.~{Madhusudhan}, \emph{Mon. Notices Royal Astron. Soc.},
  2016, \textbf{458}, 1330--1344\relax
\mciteBstWouldAddEndPuncttrue
\mciteSetBstMidEndSepPunct{\mcitedefaultmidpunct}
{\mcitedefaultendpunct}{\mcitedefaultseppunct}\relax
\EndOfBibitem
\bibitem[{Seager} \emph{et~al.}(2007){Seager}\emph{et~al.}]{Seager2007}
S.~{Seager} \emph{et~al.}, \emph{Astrophys. J.}, 2007, \textbf{669},
  1279--1297\relax
\mciteBstWouldAddEndPuncttrue
\mciteSetBstMidEndSepPunct{\mcitedefaultmidpunct}
{\mcitedefaultendpunct}{\mcitedefaultseppunct}\relax
\EndOfBibitem
\bibitem[Birch(1952)]{Birch1952}
F.~Birch, \emph{J. Geophys. Res. (1896-1977)}, 1952, \textbf{57},
  227--286\relax
\mciteBstWouldAddEndPuncttrue
\mciteSetBstMidEndSepPunct{\mcitedefaultmidpunct}
{\mcitedefaultendpunct}{\mcitedefaultseppunct}\relax
\EndOfBibitem
\bibitem[Karki \emph{et~al.}(2000)Karki\emph{et~al.}]{Karki2000}
B.~B. Karki \emph{et~al.}, \emph{Phys. Rev. B}, 2000, \textbf{62},
  14750--14756\relax
\mciteBstWouldAddEndPuncttrue
\mciteSetBstMidEndSepPunct{\mcitedefaultmidpunct}
{\mcitedefaultendpunct}{\mcitedefaultseppunct}\relax
\EndOfBibitem
\bibitem[Ahrens(2000)]{Ahrens2000}
T.~J. Ahrens, \emph{Mineral Physics \& Crystallography: A Handbook of Physical
  Constants}, American Geophysical Union, Washington, DC, 2000\relax
\mciteBstWouldAddEndPuncttrue
\mciteSetBstMidEndSepPunct{\mcitedefaultmidpunct}
{\mcitedefaultendpunct}{\mcitedefaultseppunct}\relax
\EndOfBibitem
\bibitem[{Scheucher} \emph{et~al.}(2020){Scheucher}, {Wunderlich}, {Grenfell},
  {Godolt}, {Schreier}, {Kappel}, {Haus}, {Herbst}, and {Rauer}]{Scheucher2020}
M.~{Scheucher}, F.~{Wunderlich}, J.~L. {Grenfell}, M.~{Godolt}, F.~{Schreier},
  D.~{Kappel}, R.~{Haus}, K.~{Herbst} and H.~{Rauer}, \emph{The Astrophysical
  Journal}, 2020, \textbf{898}, 44\relax
\mciteBstWouldAddEndPuncttrue
\mciteSetBstMidEndSepPunct{\mcitedefaultmidpunct}
{\mcitedefaultendpunct}{\mcitedefaultseppunct}\relax
\EndOfBibitem
\bibitem[{Blain} \emph{et~al.}(2021){Blain}, {Charnay}, and
  {B{\'e}zard}]{Blain2021}
D.~{Blain}, B.~{Charnay} and B.~{B{\'e}zard}, \emph{Astron. Astrophys.}, 2021,
  \textbf{646}, A15\relax
\mciteBstWouldAddEndPuncttrue
\mciteSetBstMidEndSepPunct{\mcitedefaultmidpunct}
{\mcitedefaultendpunct}{\mcitedefaultseppunct}\relax
\EndOfBibitem
\bibitem[{Charnay} \emph{et~al.}(2021){Charnay}, {Blain}, {B{\'e}zard},
  {Leconte}, {Turbet}, and {Falco}]{Charnay2021}
B.~{Charnay}, D.~{Blain}, B.~{B{\'e}zard}, J.~{Leconte}, M.~{Turbet} and
  A.~{Falco}, \emph{Astron. Astrophys.}, 2021, \textbf{646}, A171\relax
\mciteBstWouldAddEndPuncttrue
\mciteSetBstMidEndSepPunct{\mcitedefaultmidpunct}
{\mcitedefaultendpunct}{\mcitedefaultseppunct}\relax
\EndOfBibitem
\bibitem[{Pierrehumbert}(2023)]{Pierrehumbert2023}
R.~T. {Pierrehumbert}, \emph{Astrophys. J.}, 2023, \textbf{944}, 20\relax
\mciteBstWouldAddEndPuncttrue
\mciteSetBstMidEndSepPunct{\mcitedefaultmidpunct}
{\mcitedefaultendpunct}{\mcitedefaultseppunct}\relax
\EndOfBibitem
\bibitem[{Rothschild} and {Mancinelli}(2001)]{Rothschild2001}
L.~J. {Rothschild} and R.~L. {Mancinelli}, \emph{Nature}, 2001, \textbf{409},
  1092--1101\relax
\mciteBstWouldAddEndPuncttrue
\mciteSetBstMidEndSepPunct{\mcitedefaultmidpunct}
{\mcitedefaultendpunct}{\mcitedefaultseppunct}\relax
\EndOfBibitem
\bibitem[Merino \emph{et~al.}(2019)Merino\emph{et~al.}]{Merino2019}
N.~Merino \emph{et~al.}, \emph{Front. Microbiol.}, 2019, \textbf{10}, 780\relax
\mciteBstWouldAddEndPuncttrue
\mciteSetBstMidEndSepPunct{\mcitedefaultmidpunct}
{\mcitedefaultendpunct}{\mcitedefaultseppunct}\relax
\EndOfBibitem
\bibitem[{Hardegree-Ullman}
  \emph{et~al.}(2020){Hardegree-Ullman}\emph{et~al.}]{Hardegree2020}
K.~K. {Hardegree-Ullman} \emph{et~al.}, \emph{Astrophys. J. Suppl.}, 2020,
  \textbf{247}, 28\relax
\mciteBstWouldAddEndPuncttrue
\mciteSetBstMidEndSepPunct{\mcitedefaultmidpunct}
{\mcitedefaultendpunct}{\mcitedefaultseppunct}\relax
\EndOfBibitem
\bibitem[{Tsiaras} \emph{et~al.}(2019){Tsiaras}\emph{et~al.}]{Tsiaras2019}
A.~{Tsiaras} \emph{et~al.}, \emph{Nat. Astron.}, 2019, \textbf{3},
  1086--1091\relax
\mciteBstWouldAddEndPuncttrue
\mciteSetBstMidEndSepPunct{\mcitedefaultmidpunct}
{\mcitedefaultendpunct}{\mcitedefaultseppunct}\relax
\EndOfBibitem
\bibitem[{Lodders} and {Fegley}(2002)]{Lodders2002}
K.~{Lodders} and B.~{Fegley}, \emph{Icarus}, 2002, \textbf{155}, 393--424\relax
\mciteBstWouldAddEndPuncttrue
\mciteSetBstMidEndSepPunct{\mcitedefaultmidpunct}
{\mcitedefaultendpunct}{\mcitedefaultseppunct}\relax
\EndOfBibitem
\bibitem[{Moses} \emph{et~al.}(2013){Moses}\emph{et~al.}]{Moses2013}
J.~I. {Moses} \emph{et~al.}, \emph{Astrophys. J.}, 2013, \textbf{777}, 34\relax
\mciteBstWouldAddEndPuncttrue
\mciteSetBstMidEndSepPunct{\mcitedefaultmidpunct}
{\mcitedefaultendpunct}{\mcitedefaultseppunct}\relax
\EndOfBibitem
\bibitem[Seager \emph{et~al.}(2013)Seager\emph{et~al.}]{Seager2013}
S.~Seager \emph{et~al.}, \emph{Astrophys. J.}, 2013, \textbf{777}, 95\relax
\mciteBstWouldAddEndPuncttrue
\mciteSetBstMidEndSepPunct{\mcitedefaultmidpunct}
{\mcitedefaultendpunct}{\mcitedefaultseppunct}\relax
\EndOfBibitem
\bibitem[{Ranjan} \emph{et~al.}(2022){Ranjan}\emph{et~al.}]{Ranjan2022}
S.~{Ranjan} \emph{et~al.}, \emph{Astrophys. J.}, 2022, \textbf{930}, 131\relax
\mciteBstWouldAddEndPuncttrue
\mciteSetBstMidEndSepPunct{\mcitedefaultmidpunct}
{\mcitedefaultendpunct}{\mcitedefaultseppunct}\relax
\EndOfBibitem
\bibitem[{Asplund} \emph{et~al.}(2009){Asplund}\emph{et~al.}]{Asplund2009}
M.~{Asplund} \emph{et~al.}, \emph{Ann. Rev. Astron. Astrophys.}, 2009,
  \textbf{47}, 481--522\relax
\mciteBstWouldAddEndPuncttrue
\mciteSetBstMidEndSepPunct{\mcitedefaultmidpunct}
{\mcitedefaultendpunct}{\mcitedefaultseppunct}\relax
\EndOfBibitem
\bibitem[Hoehler \emph{et~al.}(2001)Hoehler\emph{et~al.}]{Hoehler2001}
T.~M. Hoehler \emph{et~al.}, \emph{FEMS Microbiol. Ecol.}, 2001, \textbf{38},
  33--41\relax
\mciteBstWouldAddEndPuncttrue
\mciteSetBstMidEndSepPunct{\mcitedefaultmidpunct}
{\mcitedefaultendpunct}{\mcitedefaultseppunct}\relax
\EndOfBibitem
\bibitem[{Bains} \emph{et~al.}(2014){Bains}, {Seager}, and {Zsom}]{Bains2014}
W.~{Bains}, S.~{Seager} and A.~{Zsom}, \emph{Life}, 2014, \textbf{4},
  716--744\relax
\mciteBstWouldAddEndPuncttrue
\mciteSetBstMidEndSepPunct{\mcitedefaultmidpunct}
{\mcitedefaultendpunct}{\mcitedefaultseppunct}\relax
\EndOfBibitem
\bibitem[Berghuis \emph{et~al.}(2019)Berghuis\emph{et~al.}]{Berghuis2019}
B.~A. Berghuis \emph{et~al.}, \emph{Proc. Natl. Acad. Sci.}, 2019,
  \textbf{116}, 5037--5044\relax
\mciteBstWouldAddEndPuncttrue
\mciteSetBstMidEndSepPunct{\mcitedefaultmidpunct}
{\mcitedefaultendpunct}{\mcitedefaultseppunct}\relax
\EndOfBibitem
\bibitem[{Robbins} \emph{et~al.}(2016){Robbins}\emph{et~al.}]{Robbins2016}
L.~J. {Robbins} \emph{et~al.}, \emph{Earth Sci. Rev.}, 2016, \textbf{163},
  323--348\relax
\mciteBstWouldAddEndPuncttrue
\mciteSetBstMidEndSepPunct{\mcitedefaultmidpunct}
{\mcitedefaultendpunct}{\mcitedefaultseppunct}\relax
\EndOfBibitem
\bibitem[Rubin \emph{et~al.}(2019)Rubin\emph{et~al.}]{Rubin2019}
M.~Rubin \emph{et~al.}, \emph{MNRAS}, 2019, \textbf{489}, 594--607\relax
\mciteBstWouldAddEndPuncttrue
\mciteSetBstMidEndSepPunct{\mcitedefaultmidpunct}
{\mcitedefaultendpunct}{\mcitedefaultseppunct}\relax
\EndOfBibitem
\bibitem[Anbar(2008)]{Anbar2008}
A.~D. Anbar, \emph{Science}, 2008, \textbf{322}, 1481--1483\relax
\mciteBstWouldAddEndPuncttrue
\mciteSetBstMidEndSepPunct{\mcitedefaultmidpunct}
{\mcitedefaultendpunct}{\mcitedefaultseppunct}\relax
\EndOfBibitem
\bibitem[Crowe \emph{et~al.}(2014)Crowe\emph{et~al.}]{Crowe2014}
S.~A. Crowe \emph{et~al.}, \emph{Science}, 2014, \textbf{346}, 735--739\relax
\mciteBstWouldAddEndPuncttrue
\mciteSetBstMidEndSepPunct{\mcitedefaultmidpunct}
{\mcitedefaultendpunct}{\mcitedefaultseppunct}\relax
\EndOfBibitem
\bibitem[Jones \emph{et~al.}(2015)Jones\emph{et~al.}]{Jones2015}
C.~Jones \emph{et~al.}, \emph{GEOLOGY}, 2015, \textbf{43}, 135--138\relax
\mciteBstWouldAddEndPuncttrue
\mciteSetBstMidEndSepPunct{\mcitedefaultmidpunct}
{\mcitedefaultendpunct}{\mcitedefaultseppunct}\relax
\EndOfBibitem
\bibitem[Saito \emph{et~al.}(2003)Saito, Sigman, and Morel]{Saito2003}
M.~A. Saito, D.~M. Sigman and F.~M. Morel, \emph{Inorganica Chim. Acta}, 2003,
  \textbf{356}, 308--318\relax
\mciteBstWouldAddEndPuncttrue
\mciteSetBstMidEndSepPunct{\mcitedefaultmidpunct}
{\mcitedefaultendpunct}{\mcitedefaultseppunct}\relax
\EndOfBibitem
\bibitem[Anbar and Knoll(2002)]{Anbar2002}
A.~D. Anbar and A.~H. Knoll, \emph{Science}, 2002, \textbf{297},
  1137--1142\relax
\mciteBstWouldAddEndPuncttrue
\mciteSetBstMidEndSepPunct{\mcitedefaultmidpunct}
{\mcitedefaultendpunct}{\mcitedefaultseppunct}\relax
\EndOfBibitem
\bibitem[Swanner \emph{et~al.}(2020)Swanner\emph{et~al.}]{Swanner2020}
E.~D. Swanner \emph{et~al.}, \emph{Earth-Science Reviews}, 2020, \textbf{211},
  103430\relax
\mciteBstWouldAddEndPuncttrue
\mciteSetBstMidEndSepPunct{\mcitedefaultmidpunct}
{\mcitedefaultendpunct}{\mcitedefaultseppunct}\relax
\EndOfBibitem
\bibitem[Tosca \emph{et~al.}(2016)Tosca\emph{et~al.}]{Tosca2016}
N.~J. Tosca \emph{et~al.}, \emph{GSA Bulletin}, 2016, \textbf{128},
  511--530\relax
\mciteBstWouldAddEndPuncttrue
\mciteSetBstMidEndSepPunct{\mcitedefaultmidpunct}
{\mcitedefaultendpunct}{\mcitedefaultseppunct}\relax
\EndOfBibitem
\bibitem[Robbins \emph{et~al.}(2013)Robbins\emph{et~al.}]{Robbins2013}
L.~J. Robbins \emph{et~al.}, \emph{GeoBiology}, 2013, \textbf{11},
  295--306\relax
\mciteBstWouldAddEndPuncttrue
\mciteSetBstMidEndSepPunct{\mcitedefaultmidpunct}
{\mcitedefaultendpunct}{\mcitedefaultseppunct}\relax
\EndOfBibitem
\bibitem[Mukherjee and Large(2020)]{Mukherjee2020}
I.~Mukherjee and R.~R. Large, \emph{Geology}, 2020, \textbf{48},
  1018--1022\relax
\mciteBstWouldAddEndPuncttrue
\mciteSetBstMidEndSepPunct{\mcitedefaultmidpunct}
{\mcitedefaultendpunct}{\mcitedefaultseppunct}\relax
\EndOfBibitem
\bibitem[Lodders(2020)]{Lodders2020}
K.~Lodders, \emph{Oxford Research Encyclopedia of Planetary Science},
  2020\relax
\mciteBstWouldAddEndPuncttrue
\mciteSetBstMidEndSepPunct{\mcitedefaultmidpunct}
{\mcitedefaultendpunct}{\mcitedefaultseppunct}\relax
\EndOfBibitem
\bibitem[{Zahnle} \emph{et~al.}(2020){Zahnle}\emph{et~al.}]{Zahnle2020}
K.~J. {Zahnle} \emph{et~al.}, \emph{Planet. Sci. J.}, 2020, \textbf{1},
  11\relax
\mciteBstWouldAddEndPuncttrue
\mciteSetBstMidEndSepPunct{\mcitedefaultmidpunct}
{\mcitedefaultendpunct}{\mcitedefaultseppunct}\relax
\EndOfBibitem
\bibitem[{Nesvorn{\'y}} \emph{et~al.}(2023){Nesvorn{\'y}}, {Roig},
  {Vokrouhlick{\'y}}, {Bottke}, {Marchi}, {Morbidelli}, and
  {Deienno}]{Nesvorny2023}
D.~{Nesvorn{\'y}}, F.~V. {Roig}, D.~{Vokrouhlick{\'y}}, W.~F. {Bottke},
  S.~{Marchi}, A.~{Morbidelli} and R.~{Deienno}, \emph{Icarus in press}, 2023,
  \textbf{399}, 115545\relax
\mciteBstWouldAddEndPuncttrue
\mciteSetBstMidEndSepPunct{\mcitedefaultmidpunct}
{\mcitedefaultendpunct}{\mcitedefaultseppunct}\relax
\EndOfBibitem
\bibitem[Gisler \emph{et~al.}(2011)Gisler, Weaver, and Gittings]{Gisler2011}
G.~Gisler, R.~Weaver and M.~Gittings, \emph{Pure Appl. Geophys.}, 2011,
  \textbf{168}, 1187--1198\relax
\mciteBstWouldAddEndPuncttrue
\mciteSetBstMidEndSepPunct{\mcitedefaultmidpunct}
{\mcitedefaultendpunct}{\mcitedefaultseppunct}\relax
\EndOfBibitem
\bibitem[Davison and Collins(2007)]{Davison2007}
T.~Davison and G.~S. Collins, \emph{Meteorit. Planet. Sci.}, 2007, \textbf{42},
  1915--1927\relax
\mciteBstWouldAddEndPuncttrue
\mciteSetBstMidEndSepPunct{\mcitedefaultmidpunct}
{\mcitedefaultendpunct}{\mcitedefaultseppunct}\relax
\EndOfBibitem
\bibitem[Nishizawa \emph{et~al.}(2020)Nishizawa\emph{et~al.}]{Nishizawa2020}
M.~Nishizawa \emph{et~al.}, \emph{Journal of Geophysical Research (Planets)},
  2020, \textbf{125}, e06291\relax
\mciteBstWouldAddEndPuncttrue
\mciteSetBstMidEndSepPunct{\mcitedefaultmidpunct}
{\mcitedefaultendpunct}{\mcitedefaultseppunct}\relax
\EndOfBibitem
\bibitem[Carry(2012)]{Carry2012}
B.~Carry, \emph{Planetary and Space Science}, 2012, \textbf{73}, 98--118\relax
\mciteBstWouldAddEndPuncttrue
\mciteSetBstMidEndSepPunct{\mcitedefaultmidpunct}
{\mcitedefaultendpunct}{\mcitedefaultseppunct}\relax
\EndOfBibitem
\bibitem[Genda \emph{et~al.}(2017)Genda, Brasser, and Mojzsis]{Genda2017}
H.~Genda, R.~Brasser and S.~Mojzsis, \emph{Earth Planet. Sci. Lett.}, 2017,
  \textbf{480}, 25--32\relax
\mciteBstWouldAddEndPuncttrue
\mciteSetBstMidEndSepPunct{\mcitedefaultmidpunct}
{\mcitedefaultendpunct}{\mcitedefaultseppunct}\relax
\EndOfBibitem
\bibitem[Johnson and Melosh(2012)]{Johnson2012}
B.~C. Johnson and H.~J. Melosh, \emph{Icarus}, 2012, \textbf{217},
  416--430\relax
\mciteBstWouldAddEndPuncttrue
\mciteSetBstMidEndSepPunct{\mcitedefaultmidpunct}
{\mcitedefaultendpunct}{\mcitedefaultseppunct}\relax
\EndOfBibitem
\bibitem[{Reiners} and {Turchyn}(2018)]{Reiners2018}
P.~W. {Reiners} and A.~V. {Turchyn}, \emph{Geology}, 2018, \textbf{46},
  863--866\relax
\mciteBstWouldAddEndPuncttrue
\mciteSetBstMidEndSepPunct{\mcitedefaultmidpunct}
{\mcitedefaultendpunct}{\mcitedefaultseppunct}\relax
\EndOfBibitem
\bibitem[Peucker-Ehrenbrink \emph{et~al.}(2016)Peucker-Ehrenbrink, Ravizza, and
  Winckler]{Peucker2016}
B.~Peucker-Ehrenbrink, G.~Ravizza and G.~Winckler, \emph{Elements}, 2016,
  \textbf{12}, 191--196\relax
\mciteBstWouldAddEndPuncttrue
\mciteSetBstMidEndSepPunct{\mcitedefaultmidpunct}
{\mcitedefaultendpunct}{\mcitedefaultseppunct}\relax
\EndOfBibitem
\bibitem[{Fung} \emph{et~al.}(2000){Fung}\emph{et~al.}]{Fung2000}
I.~Y. {Fung} \emph{et~al.}, \emph{Global Biogeochem. Cycles}, 2000,
  \textbf{14}, 281--295\relax
\mciteBstWouldAddEndPuncttrue
\mciteSetBstMidEndSepPunct{\mcitedefaultmidpunct}
{\mcitedefaultendpunct}{\mcitedefaultseppunct}\relax
\EndOfBibitem
\bibitem[{Constantinou} and {Madhusudhan}(2022)]{Constantinou2022}
S.~{Constantinou} and N.~{Madhusudhan}, \emph{Mon. Notices Royal Astron. Soc.},
  2022, \textbf{514}, 2073--2091\relax
\mciteBstWouldAddEndPuncttrue
\mciteSetBstMidEndSepPunct{\mcitedefaultmidpunct}
{\mcitedefaultendpunct}{\mcitedefaultseppunct}\relax
\EndOfBibitem
\bibitem[{Pinhas} \emph{et~al.}(2018){Pinhas}\emph{et~al.}]{Pinhas2018}
A.~{Pinhas} \emph{et~al.}, \emph{Mon. Notices Royal Astron. Soc.}, 2018,
  \textbf{480}, 5314--5331\relax
\mciteBstWouldAddEndPuncttrue
\mciteSetBstMidEndSepPunct{\mcitedefaultmidpunct}
{\mcitedefaultendpunct}{\mcitedefaultseppunct}\relax
\EndOfBibitem
\bibitem[{Gordon} \emph{et~al.}(2022){Gordon}\emph{et~al.}]{HITRAN2020}
I.~E. {Gordon} \emph{et~al.}, \emph{J. Quant. Spectrosc. Radiat. Transf.},
  2022, \textbf{277}, 107949\relax
\mciteBstWouldAddEndPuncttrue
\mciteSetBstMidEndSepPunct{\mcitedefaultmidpunct}
{\mcitedefaultendpunct}{\mcitedefaultseppunct}\relax
\EndOfBibitem
\bibitem[Lattanzi \emph{et~al.}(2011)Lattanzi, Lauro, and
  Auwera]{C2H6-nu-12-1425}
F.~Lattanzi, C.~D. Lauro and J.~V. Auwera, \emph{J. Mol. Spec.}, 2011,
  \textbf{267}, 71--79\relax
\mciteBstWouldAddEndPuncttrue
\mciteSetBstMidEndSepPunct{\mcitedefaultmidpunct}
{\mcitedefaultendpunct}{\mcitedefaultseppunct}\relax
\EndOfBibitem
\bibitem[Harrison \emph{et~al.}(2010)Harrison, Allen, and Bernath]{H-658}
J.~J. Harrison, N.~D.~C. Allen and P.~F. Bernath, \emph{J. Quant. Spectrosc.
  Radiat. Transf.}, 2010, \textbf{111}, 357--363\relax
\mciteBstWouldAddEndPuncttrue
\mciteSetBstMidEndSepPunct{\mcitedefaultmidpunct}
{\mcitedefaultendpunct}{\mcitedefaultseppunct}\relax
\EndOfBibitem
\bibitem[Jolly \emph{et~al.}(2007)Jolly, Benilan, and
  Fayt]{HC3N-gamma_air-1-614}
A.~Jolly, Y.~Benilan and A.~Fayt, \emph{J. Mol. Spec.}, 2007, \textbf{242},
  46--54\relax
\mciteBstWouldAddEndPuncttrue
\mciteSetBstMidEndSepPunct{\mcitedefaultmidpunct}
{\mcitedefaultendpunct}{\mcitedefaultseppunct}\relax
\EndOfBibitem
\bibitem[Xu \emph{et~al.}(2004)Xu\emph{et~al.}]{CH3OH-gamma_air-1-568}
L.~Xu \emph{et~al.}, \emph{J. Mol. Spec.}, 2004, \textbf{228}, 453--470\relax
\mciteBstWouldAddEndPuncttrue
\mciteSetBstMidEndSepPunct{\mcitedefaultmidpunct}
{\mcitedefaultendpunct}{\mcitedefaultseppunct}\relax
\EndOfBibitem
\bibitem[M\"{u}ller \emph{et~al.}(2001)M\"{u}ller\emph{et~al.}]{CH3OH-S-2-569}
H.~M\"{u}ller \emph{et~al.}, \emph{Astron. Astrophys.}, 2001, \textbf{370},
  L49--L52\relax
\mciteBstWouldAddEndPuncttrue
\mciteSetBstMidEndSepPunct{\mcitedefaultmidpunct}
{\mcitedefaultendpunct}{\mcitedefaultseppunct}\relax
\EndOfBibitem
\bibitem[{Doyon} \emph{et~al.}(2012){Doyon}\emph{et~al.}]{Doyon2012}
R.~{Doyon} \emph{et~al.}, Space Telescopes and Instrumentation 2012: Optical,
  Infrared, and Millimeter Wave, 2012, p. 84422R\relax
\mciteBstWouldAddEndPuncttrue
\mciteSetBstMidEndSepPunct{\mcitedefaultmidpunct}
{\mcitedefaultendpunct}{\mcitedefaultseppunct}\relax
\EndOfBibitem
\bibitem[{Ferruit} \emph{et~al.}(2012){Ferruit}\emph{et~al.}]{ferruit2012}
P.~{Ferruit} \emph{et~al.}, Space Telescopes and Instrumentation 2012: Optical,
  Infrared, and Millimeter Wave, 2012, p. 84422O\relax
\mciteBstWouldAddEndPuncttrue
\mciteSetBstMidEndSepPunct{\mcitedefaultmidpunct}
{\mcitedefaultendpunct}{\mcitedefaultseppunct}\relax
\EndOfBibitem
\bibitem[{Birkmann} \emph{et~al.}(2014){Birkmann}\emph{et~al.}]{Birkmann2014}
S.~M. {Birkmann} \emph{et~al.}, Space Telescopes and Instrumentation 2014:
  Optical, Infrared, and Millimeter Wave, 2014, p. 914308\relax
\mciteBstWouldAddEndPuncttrue
\mciteSetBstMidEndSepPunct{\mcitedefaultmidpunct}
{\mcitedefaultendpunct}{\mcitedefaultseppunct}\relax
\EndOfBibitem
\bibitem[Rieke \emph{et~al.}(2015)Rieke\emph{et~al.}]{Rieke_2015}
G.~H. Rieke \emph{et~al.}, \emph{Pub. Astron. Soc. Pac.}, 2015, \textbf{127},
  584\relax
\mciteBstWouldAddEndPuncttrue
\mciteSetBstMidEndSepPunct{\mcitedefaultmidpunct}
{\mcitedefaultendpunct}{\mcitedefaultseppunct}\relax
\EndOfBibitem
\bibitem[{Batalha} \emph{et~al.}(2017){Batalha}\emph{et~al.}]{Batalha2017}
N.~E. {Batalha} \emph{et~al.}, \emph{Pub. Astron. Soc. Pac.}, 2017,
  \textbf{129}, 064501\relax
\mciteBstWouldAddEndPuncttrue
\mciteSetBstMidEndSepPunct{\mcitedefaultmidpunct}
{\mcitedefaultendpunct}{\mcitedefaultseppunct}\relax
\EndOfBibitem
\bibitem[{Meadows} and {Barnes}(2018)]{Meadows2018}
V.~S. {Meadows} and R.~K. {Barnes}, \emph{Handbook of Exoplanets}, 2018,
  p.~57\relax
\mciteBstWouldAddEndPuncttrue
\mciteSetBstMidEndSepPunct{\mcitedefaultmidpunct}
{\mcitedefaultendpunct}{\mcitedefaultseppunct}\relax
\EndOfBibitem
\bibitem[{Seager} \emph{et~al.}(2020){Seager}\emph{et~al.}]{Seager2020}
S.~{Seager} \emph{et~al.}, \emph{Nat. Astron.}, 2020, \textbf{4},
  802--806\relax
\mciteBstWouldAddEndPuncttrue
\mciteSetBstMidEndSepPunct{\mcitedefaultmidpunct}
{\mcitedefaultendpunct}{\mcitedefaultseppunct}\relax
\EndOfBibitem
\bibitem[{Kite} \emph{et~al.}(2020){Kite}\emph{et~al.}]{Kite2020}
E.~S. {Kite} \emph{et~al.}, \emph{Astrophys. J.}, 2020, \textbf{891}, 111\relax
\mciteBstWouldAddEndPuncttrue
\mciteSetBstMidEndSepPunct{\mcitedefaultmidpunct}
{\mcitedefaultendpunct}{\mcitedefaultseppunct}\relax
\EndOfBibitem
\bibitem[{Misener} and {Schlichting}(2023)]{Misener2023}
W.~{Misener} and H.~E. {Schlichting}, \emph{arXiv e-prints}, 2023,
  arXiv:2303.09653\relax
\mciteBstWouldAddEndPuncttrue
\mciteSetBstMidEndSepPunct{\mcitedefaultmidpunct}
{\mcitedefaultendpunct}{\mcitedefaultseppunct}\relax
\EndOfBibitem
\bibitem[{Madhusudhan}
  \emph{et~al.}(2016){Madhusudhan}\emph{et~al.}]{Madhu2016}
N.~{Madhusudhan} \emph{et~al.}, \emph{Space Sci. Rev.}, 2016, \textbf{205},
  285--348\relax
\mciteBstWouldAddEndPuncttrue
\mciteSetBstMidEndSepPunct{\mcitedefaultmidpunct}
{\mcitedefaultendpunct}{\mcitedefaultseppunct}\relax
\EndOfBibitem
\end{mcitethebibliography}
\bibliographystyle{rsc} 

\end{document}